\documentclass[preprint]{aastex}
\newcommand{\asec}	{\mbox{$^{\prime \prime}  $} }
\newcommand{\amin}	{\mbox{$^{\prime}$}}
\begin{document}
\title{A Deep, Wide Field, Optical, And Near Infrared Catalog Of A Large Area Around The Hubble Deep Field North$^{1}$}
\author{P. Capak\altaffilmark{2}, L.L. Cowie\altaffilmark{2}, E.M. Hu\altaffilmark{2}, A.J. Barger\altaffilmark{2,3,4}, M. Dickinson\altaffilmark{5}, E. Fernandez\altaffilmark{6}, M. Giavalisco\altaffilmark{5}, Y. Komiyama\altaffilmark{7}, C. Kretchmer\altaffilmark{8}, C. McNally\altaffilmark{9}, S. Miyazaki\altaffilmark{7}, S. Okamura\altaffilmark{10}, D. Stern\altaffilmark{11,12}}

\altaffiltext{1}{Based in part on data collected at Subaru Telescope, which is operated by the National Astronomical Observatory of Japan.}
\altaffiltext{2}{University of Hawaii, Institute for Astronomy, 2680 Woodlawn Drive, Honolulu, HI, 96822, USA}
\altaffiltext{3}{Department of Physics and Astronomy, University of Hawaii, 2505 Correa Road, Honolulu, HI, 96822, USA}
\altaffiltext{4}{Department of Astronomy, University of Wisconsin-Madison, 475 North Charter Street, Madison, WI, 53706, USA}
\altaffiltext{5}{Space Telescope Science Institute, Baltimore, MD, 21218, USA}
\altaffiltext{6}{New Mexico Institute of Mining and Technology, 801 Leroy Place, Socorro, NM, 87801, USA}
\altaffiltext{7}{Subaru Telescope, 650 North A'ohoku Place, Hilo, Hawaii 96720, USA}
\altaffiltext{8}{Department of Physics and Astronomy, Johns Hopkins University, 3400 North Charles Street, Baltimore, MD, 21218-2686, USA}
\altaffiltext{9}{University of Maine, Department of Physics and Astronomy, Orono,ME, 04469-5709, USA}
\altaffiltext{10}{Department of Astronomy and Research Center for the Early Universe, School of Science, University of Tokyo, Bunkyo-ku, Tokyo, 113-0033 Japan}
\altaffiltext{11}{Jet Propulsion Laboratory, California Institute of Technology, Mail Stop 169-327, Pasadena, CA, 91109, USA}
\altaffiltext{12}{Visiting Astronomer, Kitt Peak National Observatory, National Optical Astronomy Observatory, which is operated by the Association of Universities for Research in Astronomy, Inc. (AURA) under cooperative agreement with the National Science Foundation.}

\begin{abstract}
We have conducted a deep multi-color imaging survey of 0.2 degrees$^{2}$ centered on the Hubble Deep Field North (HDF-N). We shall refer to this region as the Hawaii-HDF-N. Deep data were collected in $U$, $B$, $V$, $R$, $I$, and $z^{\prime}$ bands over the central 0.2 degrees$^{2}$ and in $HK^{\prime}$ over a smaller region covering the Chandra Deep Field North (CDF-N). The data were reduced to have accurate relative photometry and astrometry across the entire field to facilitate photometric redshifts and spectroscopic followup.  We have compiled a catalog of 48,858 objects in the central 0.2 degrees$^{2}$ detected at $5\sigma$ significance in a 3\asec aperture in either $R$ or $z^{\prime}$ band. Number counts and color-magnitude diagrams are presented and shown to be consistent with previous observations.  Using color selection we have measured the density of objects at $3<z<7$.  Our multi-color data indicates that samples selected at $z > 5.5$ using the Lyman break technique suffer from more contamination by low redshift objects than suggested by previous studies.
\end{abstract}


\section{Introduction}
Deep surveys provide numerous constraints on the structure and evolution of the universe.  It has been known that galaxy number counts in a specific bandpass can provide a useful constraint on galaxy evolution \citep{1972ApJ...173L..93T,1974ApJ...194..555B,1980ApJ...241...41T}.  Multi-color surveys further constrain the galaxy formation history and allow for the selection of galaxies in specific redshift ranges.  This work was pioneered in the Hawaii surveys \citep{1988prun.proc....1C} and subsequently used extensively by Steidel et. al \citep{1996AJ....112..352S,1999ApJ...519....1S} for mapping $z \simeq 3$ and $z \simeq 4$ galaxies.  By using a color selection these groups were able to select and obtain spectra for over 1000 $z \simeq 3$ galaxies with a 90\% success rate.  This gave us a wealth of information about this redshift range.  With the ability of modern detectors to take deep images in the near IR we can now select galaxies up to $z > 6.5$ \citep{2002ApJ...568L..75H}.  The Hubble Deep Fields (HDFs) \citep{1996AJ....112.1335W} showed the value of photometric redshifts applied to deep multi-color surveys.  This technique allows estimates of redshifts for objects much too faint for optical spectroscopy.  It allows for economical measurement of the redshifts of millions of galaxies. 

Despite the depth and accuracy of the HDF data, it only covers a small co-moving volume. \citet{1998ApJ...506...33P} have looked at the clustering of galaxies using the two point correlation function of galaxies projected on the sky.  This work covered 16 sq. degrees in $I$ band.  It has shown that galaxies are highly clustered on scales of 15-20 $h^{-1}$ Mpc where $h=$ Ho/100 km s$^{-1}$Mpc$^{-1}$. This has provided a challenge in obtaining an unbiased sample of galaxies, since several square degrees must be surveyed. The present paper is the first of a series which will aim to obtain the required data by imaging several 0.2 sq. degree fields.

Previous studies of galaxy formation and evolution have solely used optical data which introduced biases.  However new generations of radio, X-ray, and mid-IR telescopes have opened up these wavelengths to deep wide-field surveys.  X-rays have constrained the Active Galactic Nuclei (AGN) history \citep{2002AJ....124.1839B}, while radio and mid-IR fluxes have proven to be a good probe of star formation.  Space based optical imaging also allowed for the morphological study of galaxies out to high redshift.  As such we have chosen to target regions of the sky that have been observed in multiple wave bands.  We began with the HDF-N because of its deep and complete multi-wavelength data provided by the Great Observatories Origins Deep Survey (GOODS)\footnote{http://www.stsci.edu/ftp/science/goods/} and other surveys.  We have covered 0.4 square degrees centered on the HDF-N.  The central 0.2 square degrees of this data are of consistent quality in all optical bands, this area contains 48,858 objects detected at 5$\sigma$ in $R$ or $z^{\prime}$ band.

This paper is one in a series focusing on the Hawaii HDF-N.  The properties of X-ray selected objects are discussed in \citet{Barger2003,2002AJ....124.1839B}. Photometric redshifts, spectroscopic redshifts, and multi-wavelength analysis will be discussed in \citet{Capak2003}. The main focus of the present paper is the optical/IR data reduction and catalog.  We also provide number counts and color-color plots to allow for comparison with current and future surveys.  We have also selected high redshift candidates using color selections and comment on the effectiveness of this technique at $z>5$ as well as the implications for the star formation history.

\section{Data Reduction}
To optimize our data collection, observations were conducted on a variety of instruments.  The $U$ band data were collected using the Kitt Peak National Observatory 4m (KPNO-4m) telescope with the MOSAIC prime focus camera.  This camera has a reasonable $U$ band response and $36\amin \times 36\amin$ field of view \citep{1998SPIE.3355..721J,1998SPIE.3355..577M,1998SPIE.3355..487W}.  The $B$, $V$, $R$, $I$, and $z^{\prime}$ band data were collected using the Subaru 8.2m telescope and Suprime-Cam instrument \citep{2002PASJ...54..833M} which is optimized for red optical response and has a $34\amin \times 27\amin$ field of view.  Finally our $HK^{\prime}$ data were collected using the QUIRC camera on the University of Hawaii 2.2m telescope \citep{1996NewA....1..177H} with a $3.6\amin \times 3.6\amin$ field of view.  The $HK^{\prime}$ filter covers both the $H$ and $K^{\prime}$ bands in a single filter which allows for greater depth at the expense of some color information. The details of these observations are found in tables \ref{data details1} and  \ref{data details2}.  

\begin{figure}
\plotone{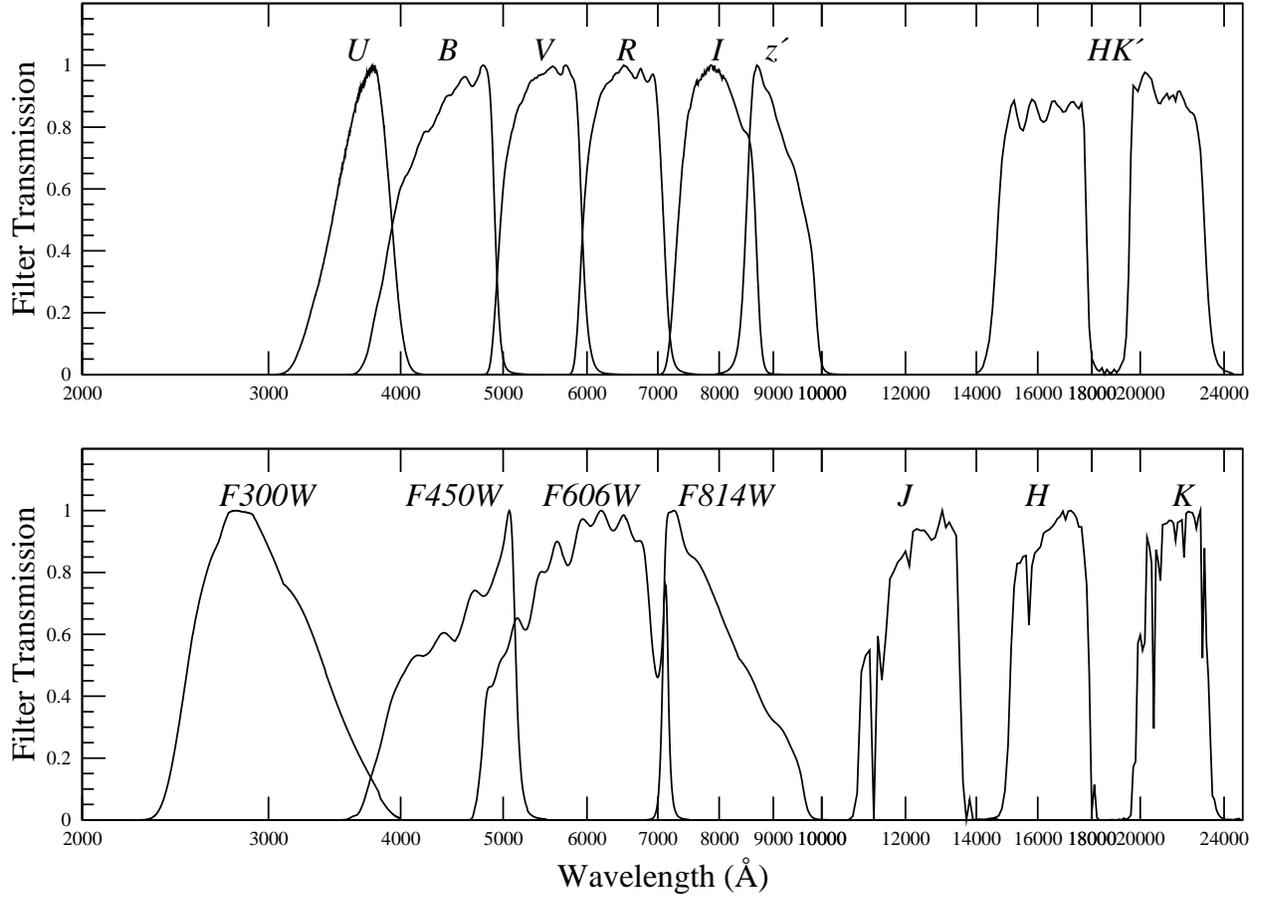}
\caption{Filter profiles multiplied by the quantum efficiency of the detectors used from our survey (top) and the HDF-N proper \citep{1999ApJ...513...34F} (bottom).  All profiles were normalized to have a peak transmission of one.\label{filters}}
\end{figure}

For data collected with Suprime-Cam a five point X-shaped dither pattern was used with one arc minute steps.  The camera was rotated by 90 degrees between dither patterns to remove bleeding from bright stars and provide better photometric calibration.  The KPNO-4m data were collected using a nine point grid as a dither pattern with one arc minute steps between pointings.  For both the Suprime-Cam and KPNO-4m observations the telescope was offset by 10\asec in a random direction between dither patterns. The $HK^{\prime}$ data were collected using a 13 point diamond shaped dither pattern with 10\asec steps between exposures.  The filter profiles multiplied by the detector response are plotted in Figure \ref{filters} along with the HDF-N filters used in \citet{1999ApJ...513...34F}.  Quantifiable measures of image quality are given in Section \ref{s:detect}.
 
\subsection{Flat Fielding}
All of these data were reduced using Nick Kaiser's Imcat tools\footnote{http://www.ifa.hawaii.edu/$\sim$kaiser/imcat/}.  The $U$, $B$, $V$, $R$, $I$, and $z^{\prime}$ data were first overscan corrected and bias subtracted.  Median sky flats were then used to provide a first pass flat fielding.  Objects were then masked out and a second flat field was generated for each dither pattern.  The second pass flat fielding was necessary to correct for changes in the instruments due to mechanical flexure and rotation of the camera and optics.  The sky was then subtracted using a second order polynomial surface fit to the sky for each chip of each image.  After the first pass sky subtraction the Suprime-Cam images had time variable scattered light structure left in the images.  This problem occurred at the edges of the field where the camera was vignetted and was worst when the moon was up.  To correct for this structure the objects were masked out and a surface was tessellated to the sky background and subtracted off.  

The $z^{\prime}$ band data taken before April of 2001 were corrected for fringing.  To achieve this a surface was tessellated on a 32-pixel grid over the first pass flat field.  This surface was used to provide first pass flat fielding instead of the median flat.  A median fringe frame was then generated for each night of data, scaled to the background in each image, and subtracted from the flat-fielded images. The images were then second pass flattened and sky subtracted as with the other bands.  After April of 2001 new chips were installed that had minimal fringing so fringe subtraction was not performed.  

The $U$ band data were corrected for internal pupil reflections which occur on the MOSAIC camera on the KPNO-4m telescope.  A pupil image was constructed by dividing the $U$ band flat by a $V$ band flat and masking out the areas that did not contain a pupil reflection.  The pupil image was subtracted from the $U$ band flat before flat fielding.  The pupil image was then scaled to the background in each image and subtracted off.  A second pass flat fielding was then performed for each dither pattern as done for the other bands.

The $HK^{\prime}$ data were collected in 13 point dither patterns so the data could be flattened and sky subtracted separately for each dither pattern using median sky flats.  This allowed us to remove the time variability in the bias, flat field, and sky which occurs in IR arrays.  The images in each dither pattern were then combined using a weighted mean to provide an image deep enough to map onto the optical astrometry grid.  Cosmic rays were removed during the combination process by looking for pixels which lay more than 5$\sigma$ from the weighted mean value.  Sigma was calculated from the background noise in each image.  Since the image depth varied across the combined, dithered images, an inverse variance map was created for each dither sequence.  These inverse variance maps were later used to combine the separate dither sequences.

\subsection{Astrometry \label{s:astrom}}
An initial solution for the optical distortion was calculated by comparing standard star frames to the USNO-A2.0 \citep{1998yCat.1252....0M}. From this point onward we treated each chip in each exposure as a separate image with a separate distortion.  This took into account changes in the optical distortion with time, as well as any mechanical motion on the focal plane.  

Using the initial solution we registered images to one another to cross identify point sources from image to image.  We also identified any USNO-A2.0 \citep{1998yCat.1252....0M} stars which were not saturated in our images.  Since most USNO-A2.0 objects were saturated in the Suprime-Cam images, the MOSAIC $U$ band data were used for the astrometric solution.  The $U$ band data was also chosen because it covers a larger area than the other bands.  

At magnitudes fainter than 18 the USNO-A2.0 stars are not properly corrected for magnitude dependent systematics.  As well errors for individual stars are not quoted in the USNO-A2.0 and are significantly worse than the mean at the faintest level.  As such all USNO-A2.0 stars were assumed to have a Root Mean Square (RMS) uncertainty of 2\asec in relative astrometry which is larger than any expected systematics. 

In the center of the field USNO-A2.0 stars were only used to constrain the scale factor and absolute astrometry of the fit. The relative astrometry was calculated by minimizing the positional scatter of stars which appeared in multiple images.  This was done by fitting a third order two-dimensional polynomial to each image.  The coefficients of the polynomials were calculated by minimizing the chi-squared defined by equation \ref{astromeq} with respect to the coefficients defined in $f(x,y)$, the polynomial function. In equation \ref{astromeq} $N_{i}$ is the number of images, and $N_{s}$ is the number of stars used for the astrometry.  In this equation the USNO-A2.0 is treated as an image, however $f(x,y)$ is replaced with the known star positions.  This effectively constrains the scale factor and rotation of the image.  Stars which were incorrectly cross-identified between images were removed in an iterative fashion until the astrometric solutions converged.

\begin{equation}
	\chi^{2} = \sum_{i=0}^{N_{i}}\sum_{j \ne i}^{N_{i}}\sum_{s=0}^{N_{s}}\frac{[f_{i}(x_{s},y_{s})-f_{j}(x_{s},y_{s})]^{2}}{\sigma_{i}^{2}+\sigma_{j}^{2}}
\label{astromeq}
\end{equation}

The resulting relative astrometry should be good to the centroiding error of $\approx0.1$ pixel or $\approx0.03$\asec RMS across the center of the field where there are many measurements for each source.  However this method breaks down at the edge of the field where there are fewer images.  At the very edge of the field the astrometry is only good to 0.5\asec since only the USNO-A2.0 constrains the fit hence systematics can not be removed.  

Any chromatic aberrations in the astrometry should be minimal since both the KPNO-4m and Suprime-Cam have an atmospheric dispersion corrector.  A large number of objects, with a wide range of colors were used for the astrometry, so any residual effects should be averaged out.  A more extensive discussion of this issue can be found in \cite{2000ApJ...537..555K}.

Once an astrometric grid of objects was established, the $B$, $V$, and, $R$ data were warped onto it using a two-dimensional, third order polynomials for each image (still treating each chip separately).  To improve the grid in the redder bands point sources from the astrometricaly calibrated $R$ band image were added to the $U$ band astrometric grid.  The $I$, $z^{\prime}$, and $HK^{\prime}$ data were warped onto this improved grid in the same way as the bluer data.  To provide accurate absolute astrometry the final images were registered to the radio catalog of \citet{2000ApJ...533..611R}.  

\begin{figure}
\begin{center}
\begin{tabular}{cc}
\includegraphics[scale=0.4]{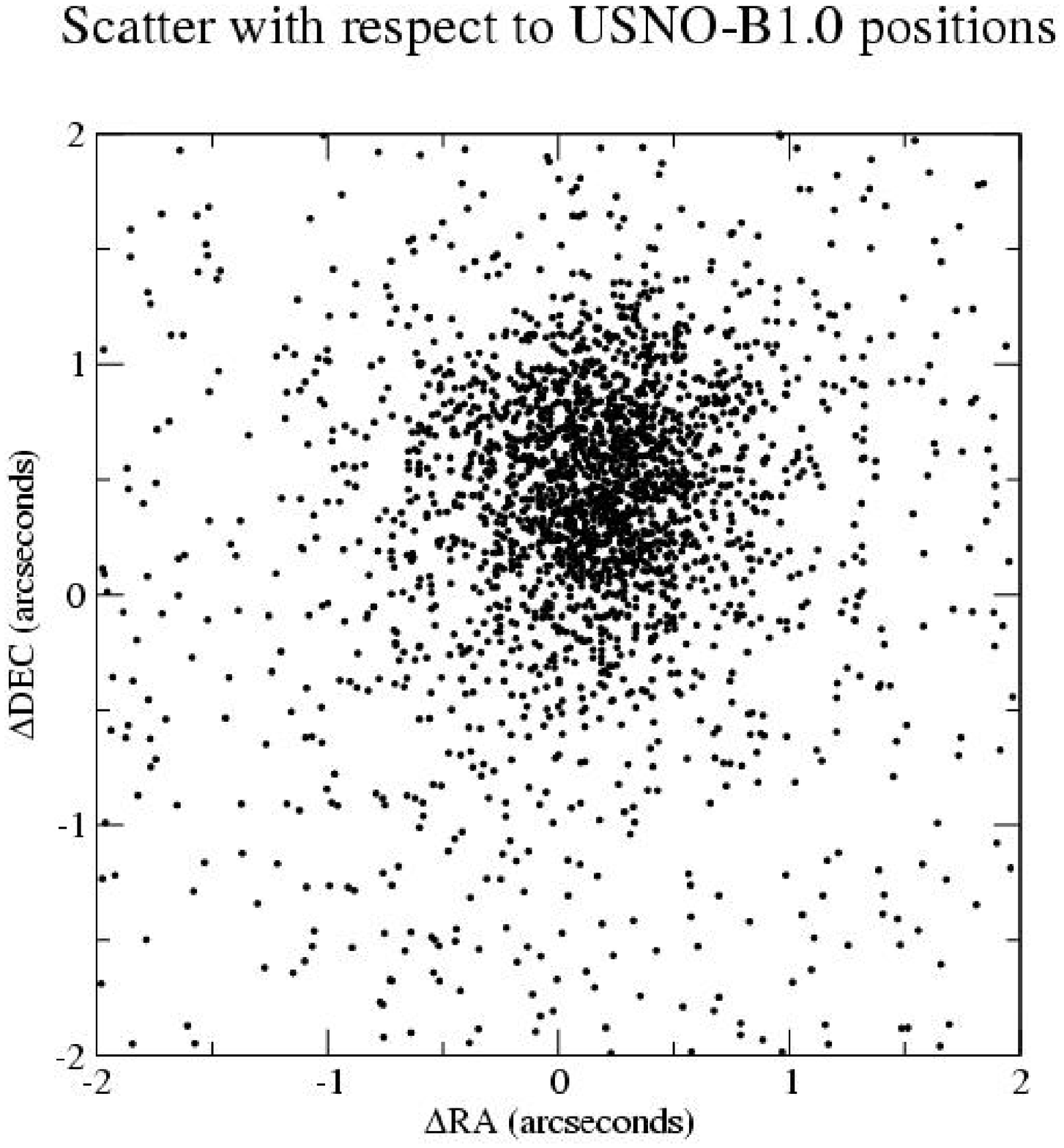}&\includegraphics[scale=0.4]{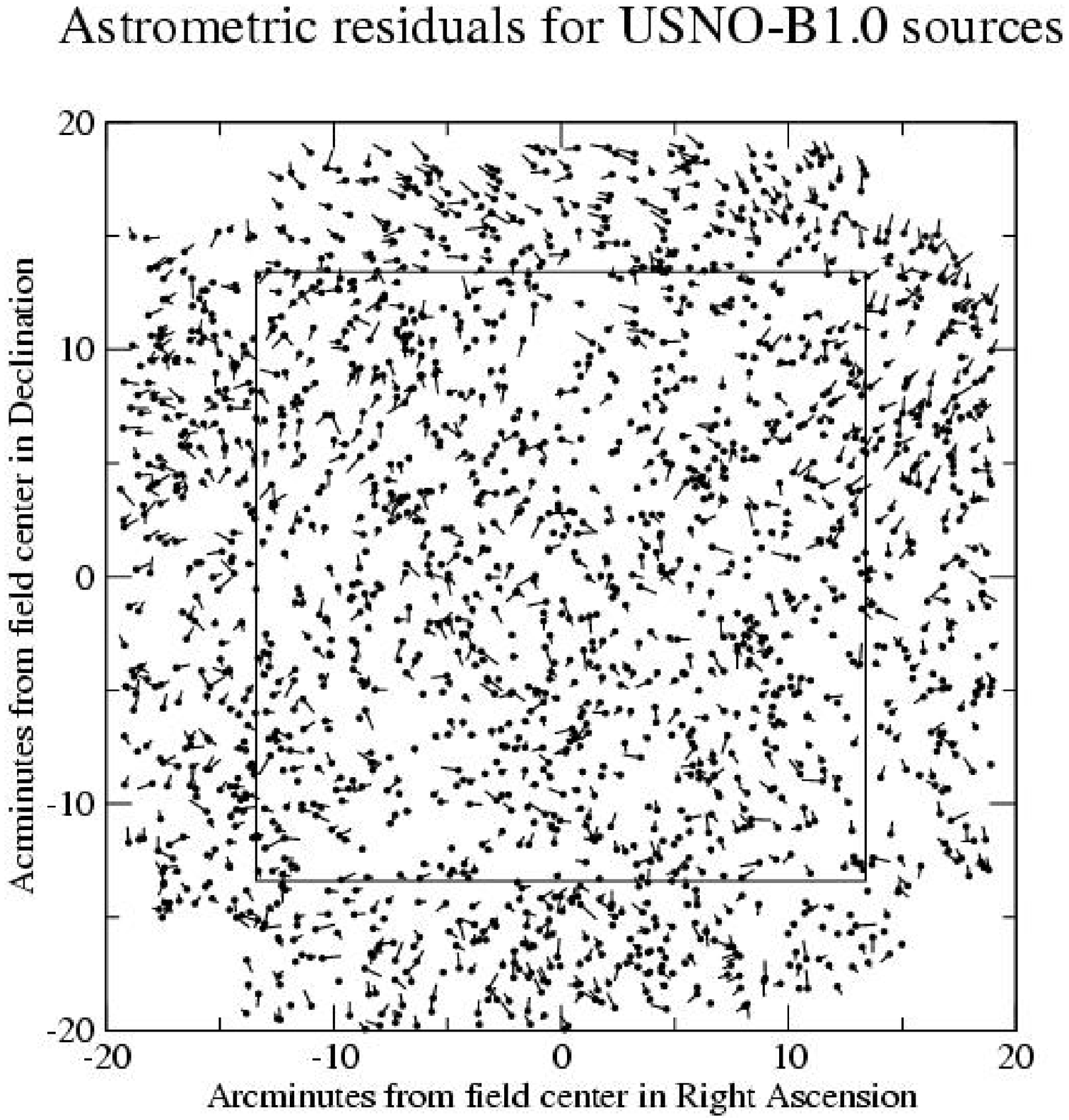}\\
\end{tabular}
\end{center}
\caption{Astrometric residuals with respect to the USNO-B1.0 \citep{2003AJ....125..984M}. The inner box on the left hand side marks the central 0.2 square degrees in the residual map. The vectors point in the direction of the residual and have and amplitude of 60 times the residual in arc seconds.\label{astro-usnobfig}}
\end{figure}

We have included comparisons to the radio catalog and USNO-B1.0 \citep{2003AJ....125..984M} in Figures \ref{astro-radiofig}, and \ref{astro-usnobfig}.  The USNO-B1.0 positions are systematically offset from the radio positions by 0.5\asec north and 0.2\asec east (see Figure \ref{astro-usnobfig}), but this offset was removed before making other comparisons. The RMS astrometric scatter between the our positions and the radio positions over the central 0.2 square degree field is 0.22\asec.  It is 0.32\asec between our positions and the USNO-B1.0. In both cases the scatter is dominated by the astrometric errors in the reference catalogs which are 0.16\asec in the radio catalog and 0.34\asec in the USNO-B1.0 and shows no systematics across the central 0.2 square degree field. As expected we observe systematics with respect to the USNO-B1.0 catalog at the edges of the field, however these offsets are less than 0.5\asec in amplitude.  Furthermore systematics at this level are known to exist in the USNO-B1.0 \citep{2003AJ....125..984M} so we chose not to remove them.

\subsection{Image Warping And Combination}

For the $U$, $B$, $V$, $R$, $I$, and $z^{\prime}$ data chip-to-chip photometric scaling factors were calculated for each dither pattern.  This was necessary because each dither pattern was flattened separately which can change the relative normalization of the chips.  When collecting our data we used large steps in our dither patterns and rotated the camera when possible.  This meant we could always find a reference chip which covered the same area of sky as two or more chips in another exposure.  The scaling between two chips in one exposure could then be calculated by comparing their scaling factors to a chip in another exposure.  This was independent of exposure to exposure scaling so long as the photometry did not vary across the reference chip.  To calculate the scale factors we used 6\asec apertures on objects detected at greater than 50$\sigma$ with half light radii less than 1.5\asec.  The large apertures were used to prevent any variations in seeing from affecting the scaling.  The large number of exposures meant we made many measurements of the chip-to-chip scaling between adjacent chips.  These were combined in a weighted mean.  Once the chip-to-chip photometric scaling was removed, exposure to exposure scale factors were calculated in a similar fashion by comparing overlapping areas between exposures. 

The geometric distortions in the MOSAIC and Suprime Cam optics cause each pixel to have a different effective area on the sky.  By flat fielding these images we introduce a geometry dependent photometric offset. When calculating the photometric scaling we corrected for this effect.  

The $HK^{\prime}$ data consists of 210 dither patterns of 13 images each.  These data were collected under a combination of photometric and non-photometric conditions.  We collected these data in such a way that each dither overlapped several neighboring dither patterns. To correct the photometry we calculated scale factors for each dither pattern such that the photometric scatter was minimized across the field in the overlapping regions.  The absolute zero point was allowed to float during this procedure and later tied to the \cite{1999ApJ...513...34F} catalog (see \S\ref{s:photom}).  

Saturated pixels were clipped in the $U$ band images before warping or combination.  However saturated pixels in the Suprime-Cam  behaved in a very non-linear fashion, often dropping or ringing in value as they became more saturated.  As such we could not simply clip the saturated pixels.  We removed the saturation spikes along with satellite trails by searching for strings of elongated objects which formed straight lines in the images.  A strip 21 pixels wide was clipped from the images along detected lines.

After applying photometric corrections each image was warped onto a stereographic projection.  We corrected the geometry dependent photometric scaling by choosing a mapping which preserved surface brightness.  The re-sampling was done using a nearest neighbor algorithm. A weighted mean was then calculated for each pixel, pixels which lay more than 5$\sigma$ from the mean were rejected to remove cosmic rays. For the $U$, $B$, $V$, $R$, $I$, and $z^{\prime}$ data the sigma used in the rejection and weighting was measured from the RMS background noise in the un-warped image. For the $HK^{\prime}$ images a sigma was calculated for each pixel during the dither pattern combination and carried through to the final image combination.  To avoid clipping pixels which varied due to seeing variations a window was defined around the median of the pixel values.  If a pixel value were between half to twice the median value of all pixels it was accepted even if it were more than 5 sigma from the mean.  A weighted mean and inverse variance were then calculated for all accepted pixels.  

\begin{figure}
\figurenum{4a}
\begin{center}
\begin{tabular}{cc}
\includegraphics[scale=0.4]{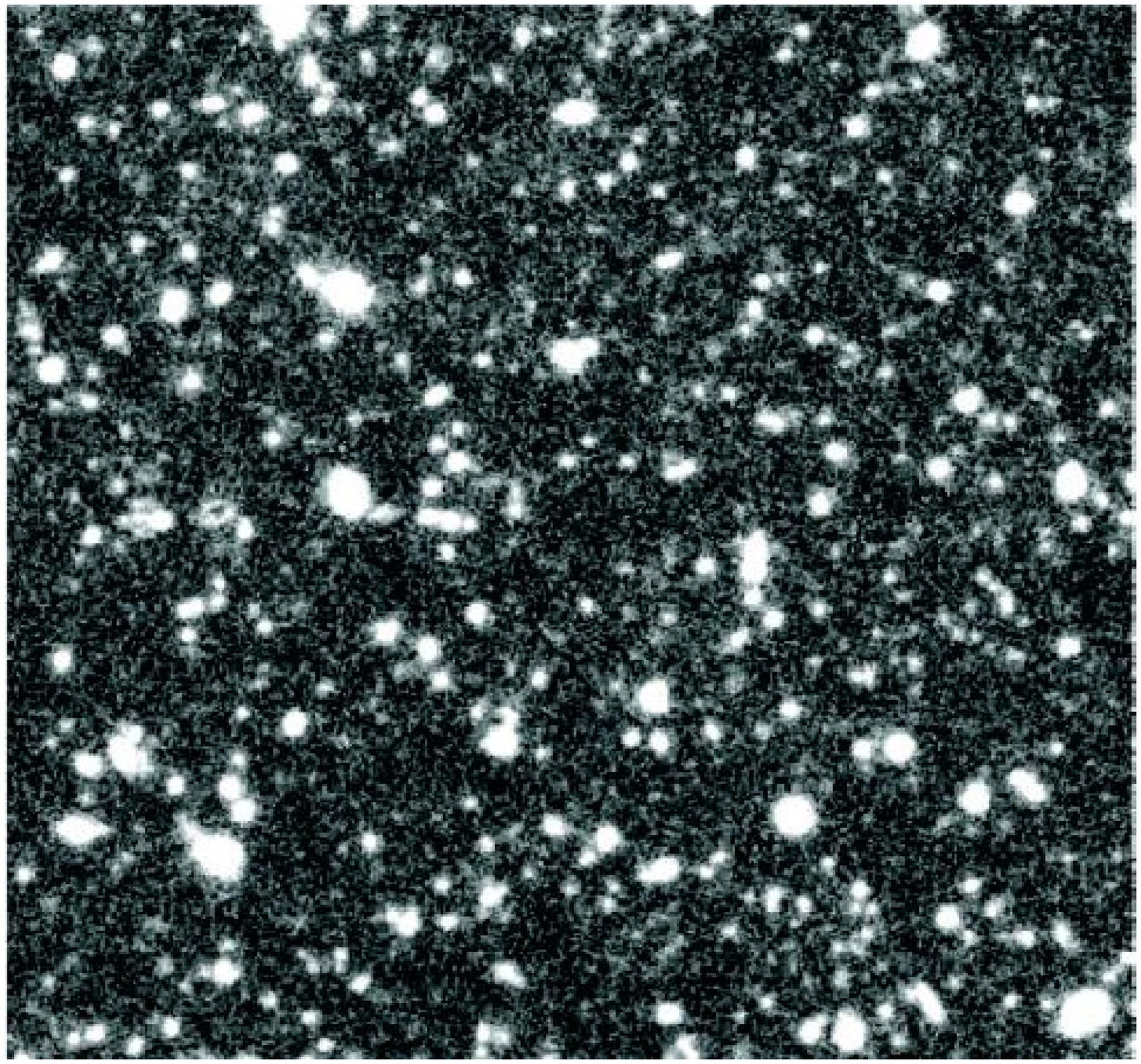}&\includegraphics[scale=0.4]{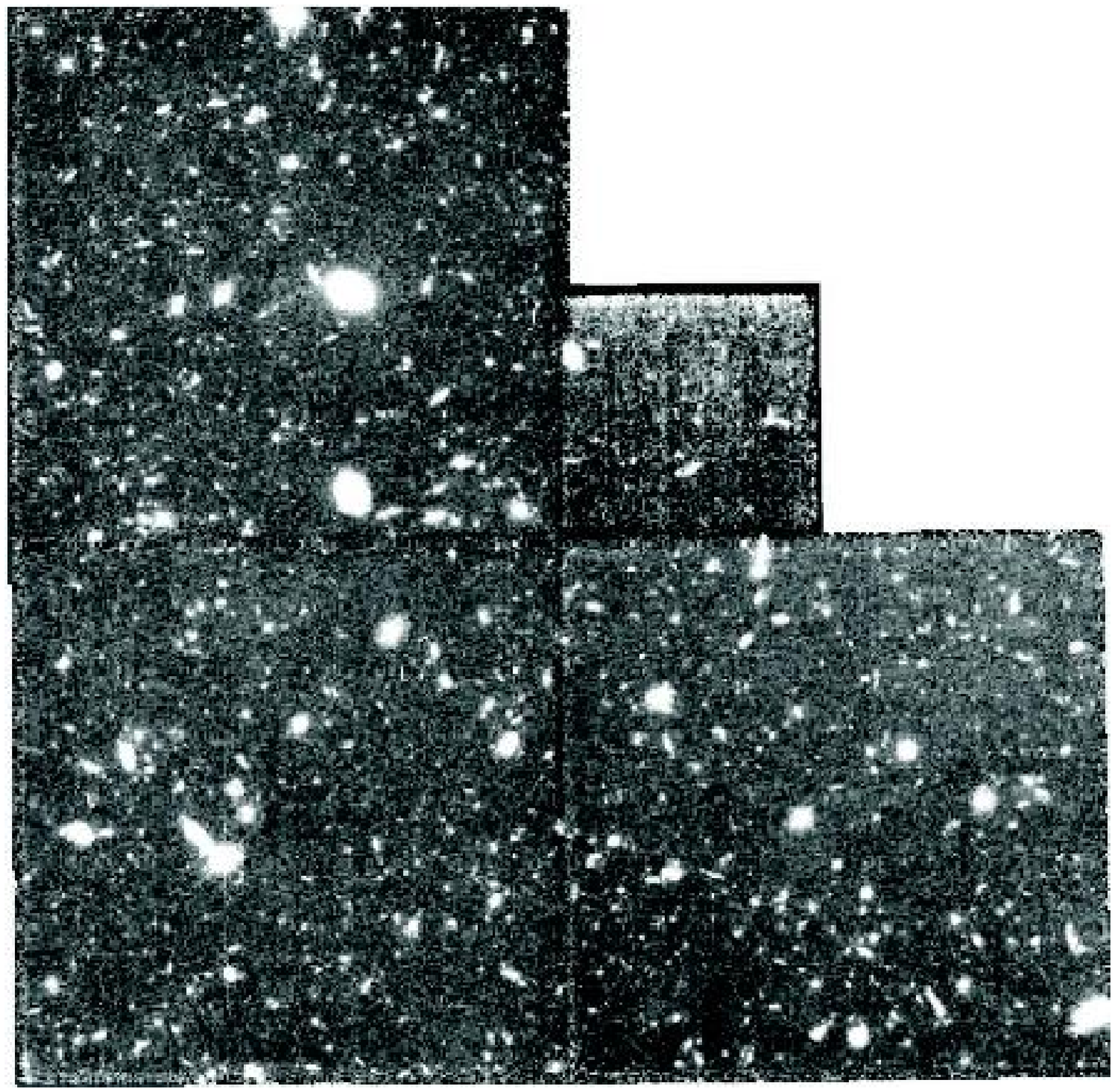}\\
\end{tabular}
\end{center}
\caption{On the left is a cutout image of our data around the HDF-N proper in $U$ band, and on the right is the Hubble space telescope $F300W$ band image for comparison\citep{1996AJ....112.1335W}.  Note the depth and quality of the ground based image.\label{Uimage}}
\end{figure}

\begin{figure}
\figurenum{4b}
\begin{center}
\begin{tabular}{cc}
\includegraphics[scale=0.4]{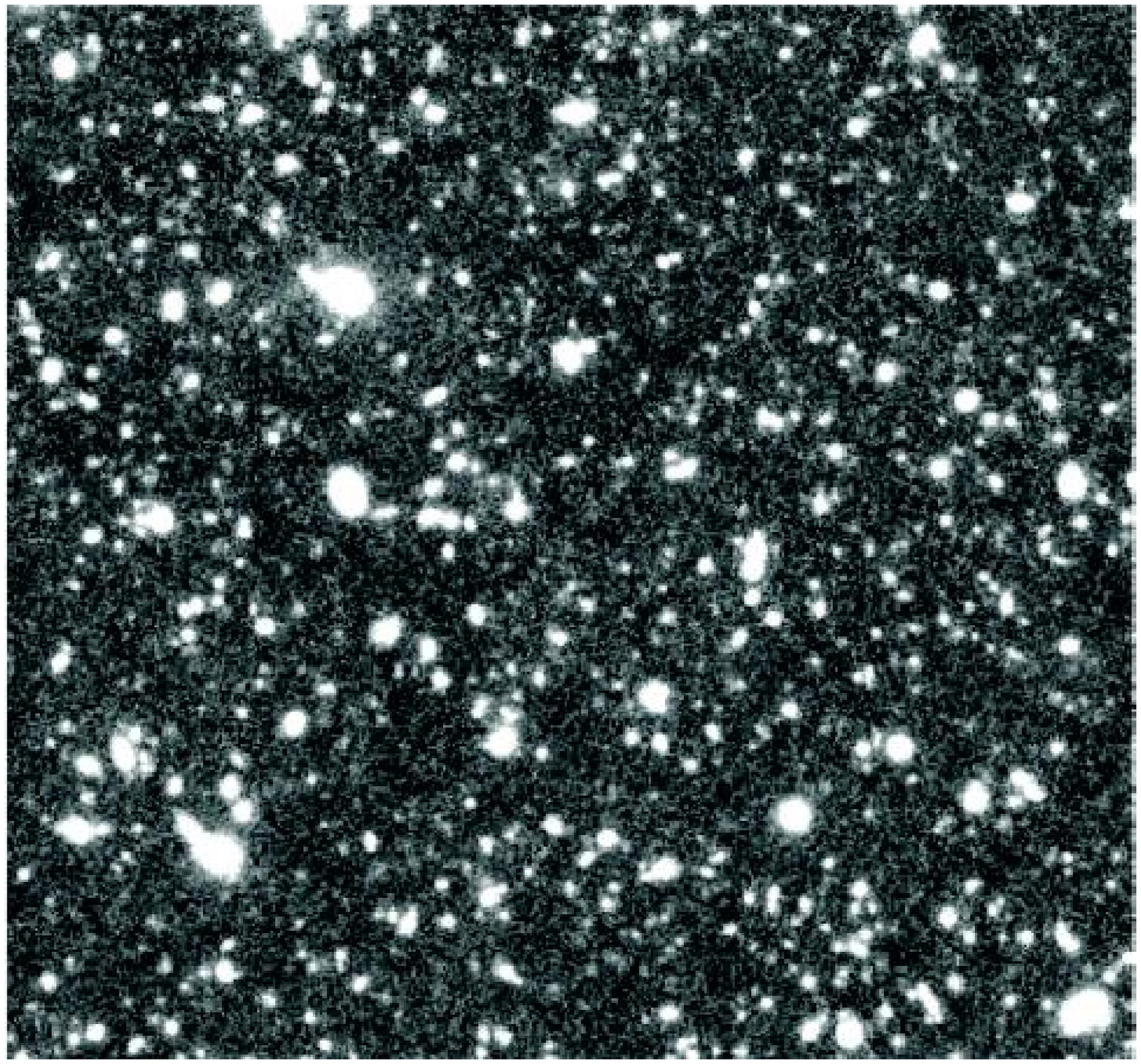}&\includegraphics[scale=0.4]{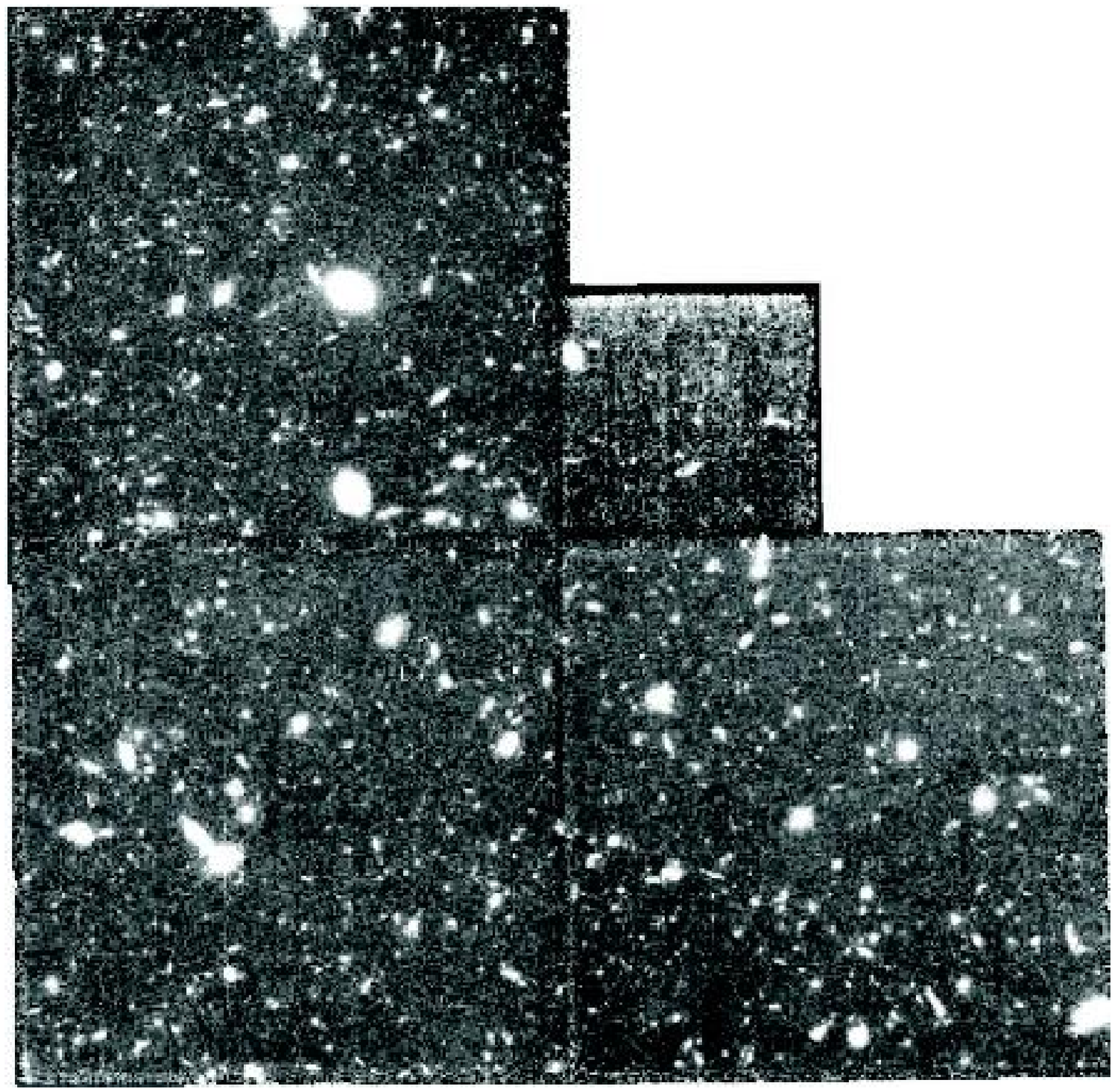}\\
\end{tabular}
\end{center}
\caption{On the left is a cutout image of our data around the HDF-N proper in $B$ band, and on the right is the Hubble space telescope $F450W$ band image for comparison\citep{1996AJ....112.1335W}.  Note the depth and quality of the ground based image.\label{Bimage}}
\end{figure}

\begin{figure}
\figurenum{4c}
\begin{center}
\begin{tabular}{cc}
\includegraphics[scale=0.4]{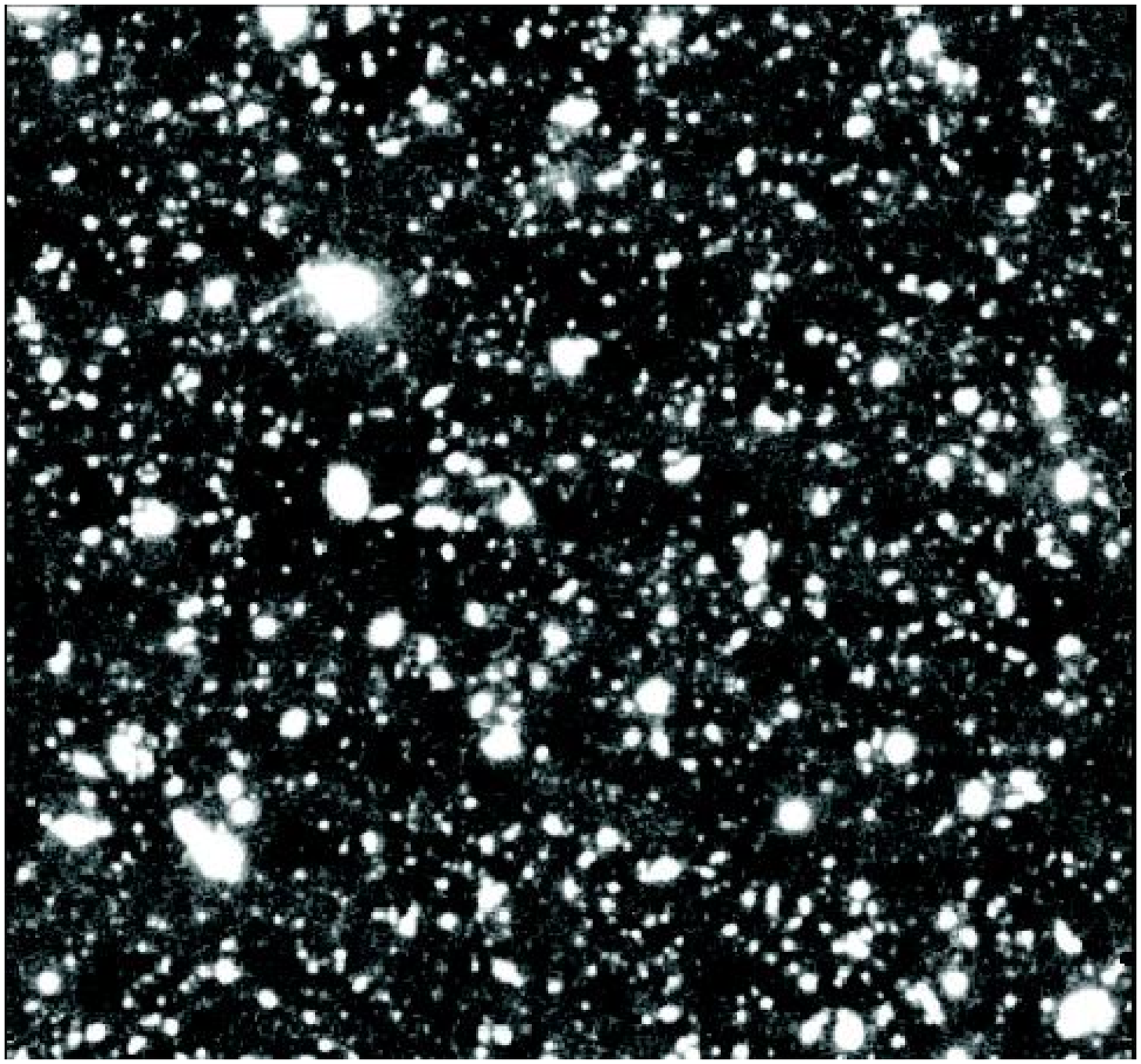}&\includegraphics[scale=0.4]{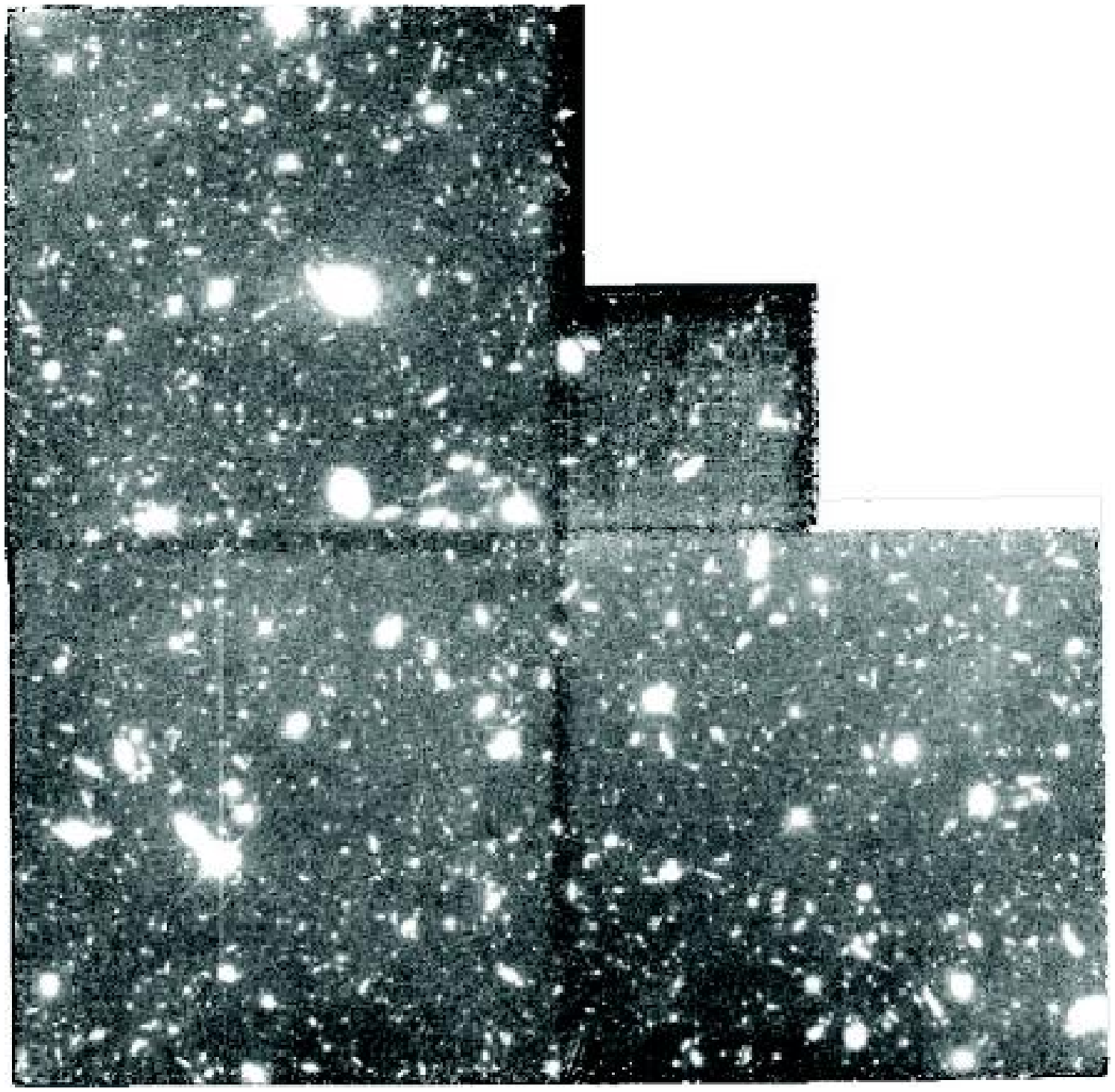}\\
\end{tabular}
\end{center}
\caption{On the left is a cutout image of our data around the HDF-N proper in $V$ band, and on the right is the Hubble space telescope $F606W$ band image for comparison\citep{1996AJ....112.1335W}.  Note the depth and quality of the ground based image.\label{Vimage}}
\end{figure}

\begin{figure}
\figurenum{4d}
\begin{center}
\begin{tabular}{cc}
\includegraphics[scale=0.4]{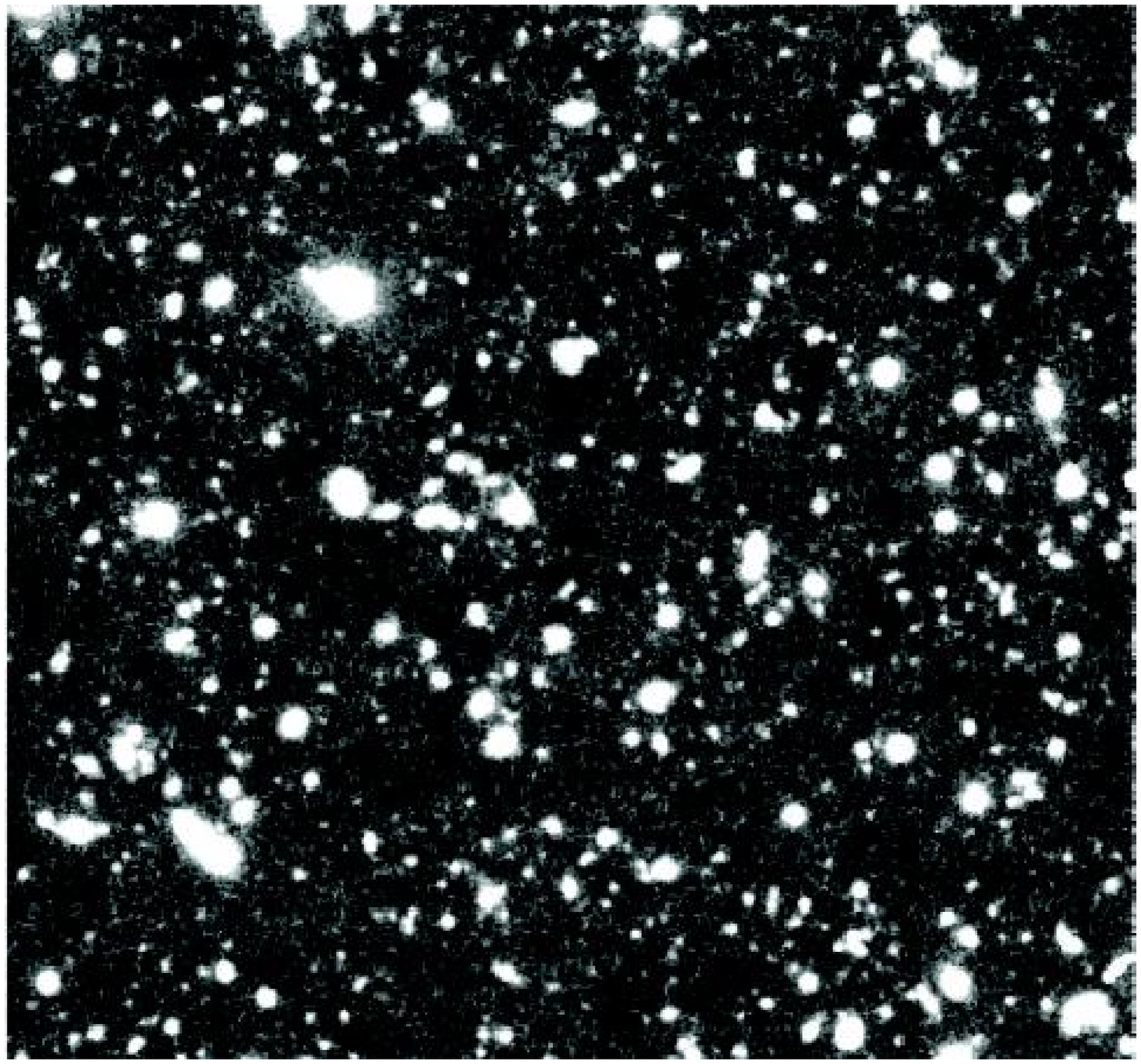}&\includegraphics[scale=0.4]{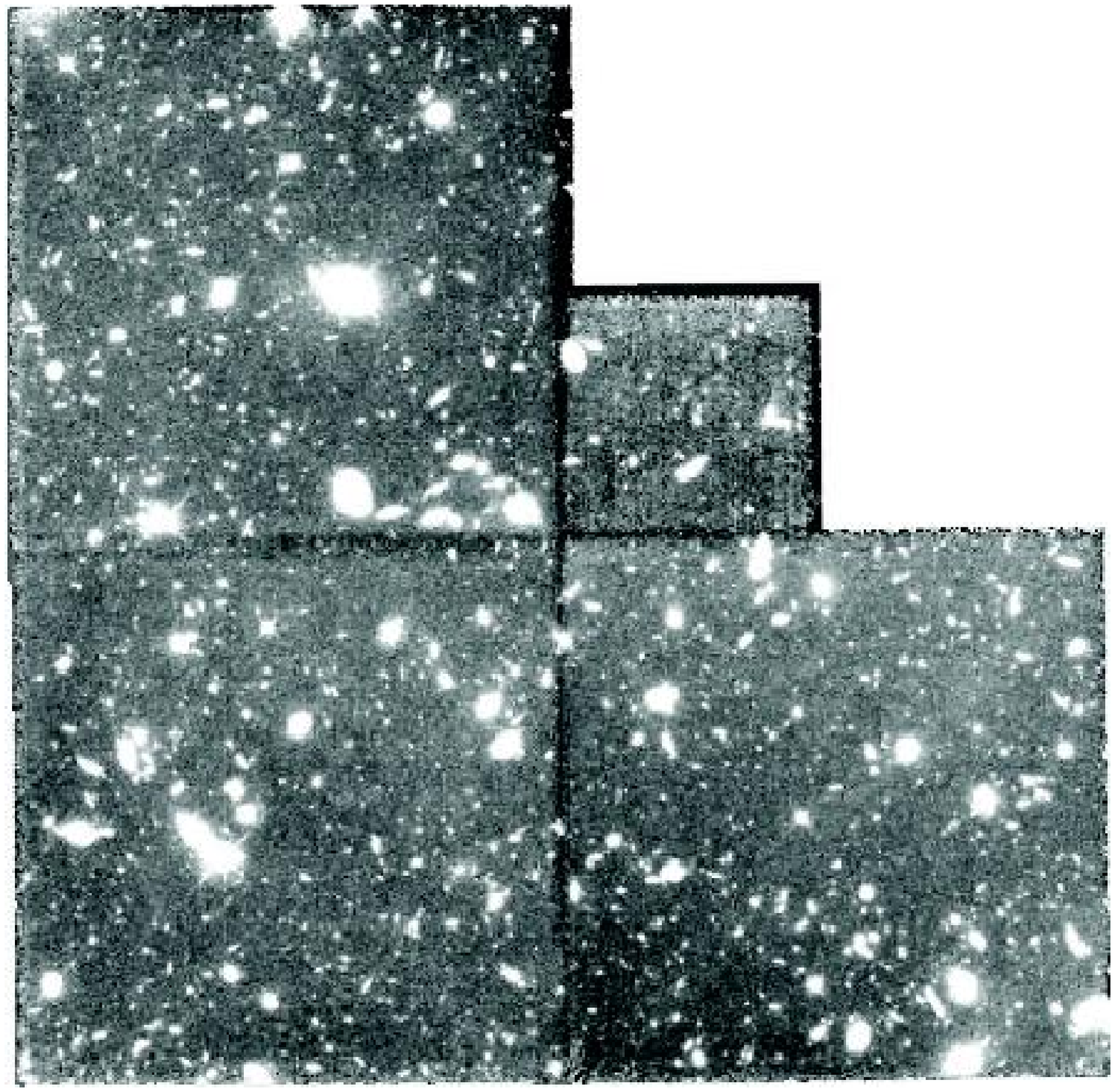}\\
\end{tabular}
\end{center}
\caption{On the left is a cutout image of our data around the HDF-N proper in $I$ band, and on the right is the Hubble space telescope $F814W$ band image for comparison\citep{1996AJ....112.1335W}. Note the depth and quality of the ground based image.\label{Iimage}}
\end{figure}

The quality of the final images is displayed in Figures \ref{Uimage}-\ref{Iimage} where we compare cutouts of our data in the HDF-N proper region with the Hubble Space Telescope images of \citet{1996AJ....112.1335W}  The area with $HK^{\prime}$ data is shown in Figure \ref{HKarea}.  

\subsection{Object Detection and Measurement \label{s:detect}}

The catalog of objects was detected in the $R$ and $z^{\prime}$ band images using Sextractor \citep{1996A&AS..117..393B}.  Objects with three or more pixels rising more than 1.5$\sigma$ above the background were analyzed, but only those with 5$\sigma$ measurements of their aperture flux were output to the catalog.  The significance of a detection was calculated using the RMS map generated during the image combination process.  The absolute scaling of the RMS map was calculated by laying down random blank apertures away from objects on the images.  This was done since the pixel-pixel noise underestimates the image noise.  This method may somewhat overestimate the true noise since it also includes variations due to faint objects in the blank apertures.  The area around bright stars was masked out to prevent erroneous detections and measurements from scattered light.

We calculated both aperture and isophotal magnitudes for our images.  Aperture magnitudes were selected as the primary magnitude to avoid biases introduced by galaxy morphology in isophotal magnitudes.  To avoid filter dependent aperture corrections the images were smoothed with a Gaussian to match the median ratio of the total to the aperture magnitude in the worst seeing ($U$ band) image. We chose this method because it ensures the apertures are sampling the same percentage of the total light in all bands and does not assume the PSF is gaussian. PSF matching provided a lower signal to noise because it includes the non-gaussian components of all seven PSF's which degraded the image quality. The aperture magnitudes were calculated on these smoothed images while the isophotal magnitudes were calculated on the un-smoothed images. A 3\asec diameter aperture was used for the photometry.  An annulus with an inner radius of 6\asec and outer radius of 12\asec was used for sky subtraction.  The 5 sigma limiting magnitude cut was calculated using aperture magnitudes on the un-smoothed image to avoid losing faint objects.

For the $V$ band data we included archival data to increase the depth \citep{2003PASJ..astroph}.  These data had better seeing (0.71\asec compared to 1.18\asec); however they did not cover our entire field.  Furthermore these data were taken with extremely long exposures which become non-linear at $\simeq$22nd magnitude  To avoid position dependent image quality and saturation, we reduced the archival data separately and scaled them to have matching photometry.  Where archival data existed and was unsaturated, it was used for the shape measurements.  Photometry was performed on the two images separately and the fluxes combined using a weighted mean.

\subsection{Absolute Photometry \label{s:photom}}

The HDF-N is a heavily observed area of the sky with extremely accurate photometry in many bands.  Many objects have known redshifts and spectral energy distributions making it easy to check the accuracy of our photometry.  Furthermore it is difficult to accurately measure standard stars with Suprime-Cam because of the size of the telescope and overheads involved.  Obtaining photometry on other telescopes would introduce problems similar to those in using the existing photometry.  As such we have chosen not to use standard stars, but rather calibrate the data using the existing photometry.

To obtain accurate absolute photometry in the AB system we matched our photometry to the HDF-N photometry of \citet{1999ApJ...513...34F} who carefully corrected for many subtle photometric effects in the HDF-N data.  We converted the fluxes in the Fern\'{a}ndez-Soto et al. catalog to our filter system by linearly interpolating the flux between the HDF-N filters.  The effective wavelength of the filters were used for the interpolation.  This method weights by transmission so any overlap between the \citet{1999ApJ...513...34F} filters and ours should be accounted for.  We then compared our isophotal fluxes to the ones we interpolated and scaled them to match.  Since the Fern\'{a}ndez-Soto et. al. fluxes were measured in an isophotal aperture, they vary systematically in magnitude with respect to the flux measured in a fixed aperture.  To prevent this from introducing filter dependent offsets in our aperture fluxes we used the same magnitude range of 22-24 $AB_{mag}$ in the $U$, $B$, $V$, $R$, $I$, and $z^{\prime}$ bands when comparing our aperture photometry to the Fern\'{a}ndez-Soto et. al. photometry.  In the $HK^{\prime}$ band we were forced to use a range of 22-23.5 because of the shallower data. This may have introduced a small bias in our photometry for this band.  We then scaled the aperture magnitudes to total magnitudes by comparing our isophotal and aperture fluxes for point sources in the $I$ band.  $I$ band was chosen because it had the best image quality of the red bands and closely approximated the $F814W$ filter used in the HDF-N.  To quantify the error in our zero point determination we measured the RMS scatter between our isophotal magnitudes and those of Fern\'{a}ndez-Soto et. al.  The RMS scatter is 0.19 magnitudes which corresponds to and average error of 0.03 magnitudes in our photometric zero points exact numbers for all bands are given in table \ref{Photometry Offsets}. Figures \ref{abs phot1} and \ref{abs phot2} compares our final photometry to the interpolated photometry of \citet{1999ApJ...513...34F}.   

Since an extensive catalog of spectroscopic redshifts \citep{2000ApJ...538...29C} exists for this field it is possible to compare our photometry to galaxy templates.  This comparison was done using the photometric redshift code of \citet{2000ApJ...536..571B} which includes a calibration mode.  First a best fit template at the known redshift was found for each object.  The templates of \citet{1980ApJS...43..393C} and \citet{1996ApJ...467...38K} were used.  Twenty intermediate spectra were interpolated between the main spectral templates to improve the fits.  Then a weighed mean offset was calculated for each filter by comparing our photometry to the template photometry.  These offsets were then applied to our photometry and the process repeated until the offsets converged.  We did not apply offsets the $B$ and $I$ magnitudes because they are close to the HDF-N $F450W$ and $F814W$ filters respectively and hence we expect them to be correctly calibrated.  The resulting offsets are small and are probably due to inaccuracies in the templates or filter profiles (see Table \ref{Photometry Offsets}).  The largest offsets occur in the $R$, $z^{\prime}$, and $HK^{\prime}$ bands which are also those least like the HDF-N filters.  The large offset in $HK^{\prime}$ may also be due to the different magnitude range used when scaling the aperture fluxes.

\subsection{Catalog Description}
Due to the large amount of data we have broken the catalog into several files which are available on the Internet\footnote{http://www.ifa.hawaii.edu/$\sim$capak/hdf/index.html}.  There are two catalogs, one selected in the $R$ band and the other in the $z^{\prime}$ band.  Objects with an aperture magnitude error smaller than 5 sigma on the un-smoothed images in the selecting band were included.  If an object was present in both catalogs it was removed from the $z^{\prime}$ catalog.  Each object has a unique ID within each catalog, but both the $R$ and $z^{\prime}$ catalog begin at 1. Our data covers up to 0.4 square degrees in some bands, however around the edges the coverage is un-even, the astrometry may be distorted (see \S\ref{s:astrom}), and the detections become unreliable due to cosmic rays, reflections and other defects which could not be removed. As such we avoided these regions for our scientific objectives.  All results in this paper are from the central 0.2 square degrees where the data is of equal depth in the $U$, $B$, $V$, $R$, $I$, and $z^{\prime}$ bands.  Since others may find it useful we have included data for the entire field in supplementary files with the same formatting and content as the main catalog. However we strongly recommend using only the main $R$ and $z^{\prime}$ catalog containing objects in the central 0.2 square degrees.  No attempt was made to ensure the integrity of the data outside the central area.

Each catalog consists of four files, the contents of which are listed in tables \ref{shape cat},\ref{flag cat},\ref{mag cat},\ref{flux cat}.  The shape file contains information on the position and morphology of the object output by Sextractor. The flag file contains flags for saturation, overlapping objects, and questionable or bad detections. The bad flag is set for objects with a Full Width at Half Max (FWHM) of 0 or with abnormal magnitudes in the selection band.  This flag was intended to mark questionable detections, however some faint point sources may also have been flagged.  The magnitude file contains aperture and isophotal magnitudes along with errors in the AB system for all bands.  We used the Sextractor convention of -99 for an object which was outside the field or which could not be measured for other reasons.  Badly saturated objects were also given a magnitude of -99.  Objects with negative fluxes were assigned negative magnitudes with the absolute value of the magnitudes corresponding to the absolute value of the flux.  The flux catalog contains aperture and isophotal fluxes along with errors and  background levels. All measurements in the flux file are in nano-Janskys (nJy).

The area around bright objects has been removed due to false detections and problematic photometry in these areas.  This was done by cutting out circles around USNO-A2.0 objects with magnitudes brighter than 15.5.  A list of these regions and the size of each cutout is provided along with the catalog.

\section{Results}
\subsection{Number Counts}

To determine number counts we constructed separate catalogs detected in each color using the un-smoothed images. We measured fluxes in 3\asec diameter apertures and applied an aperture correction to each band.  We did not use the Sextractor best or isophotal magnitudes because they often act in a non-linear way at low signal to noise.  The correction was calculated by measuring the median offset between Sextractor best magnitudes \citep{1996A&AS..117..393B}, which estimate total magnitudes, and the aperture magnitudes for objects brighter than 24th magnitude (see Table \ref{Photometry Offsets}).  Using aperture magnitudes may introduce a magnitude dependent bias in the number counts. However we saw no evidence that the median correction was changing with magnitude.  Due to the difficulty in identifying stars at faint magnitudes we have provided both raw counts and counts with stars removed.  Objects were identified as stars if they had a value of 0.9 or greater in the Sextractor star galaxy-separator.  We did not remove stars in the $HK^{\prime}$ band because the star galaxy separator was unreliable in that band due to variable seeing across the image.  The number counts shown in Figures \ref{Ucountfig}-\ref{Kcountfig} were normalized to a Euclidean slope as described in \citet{2001AJ....122.1104Y}.  They agree with those reported by other authors \citep{2001MNRAS.323..795M,2001A&A...376..756M,2001A&A...368..787H,2001AJ....122.1104Y,2001A&A...375....1S,1995ApJ...438L..13D,1994ApJ...420L...1S,1993ApJ...415L...9G}. The Sloan Digital Sky Survey counts \citep{2001AJ....122.1104Y} were measured in $u^{\prime}$, $g^{\prime}$, $r^{\prime}$, $i^{\prime}$, and $z^{\prime}$ which are close to our bands with the exception of $g^{\prime}$ and $r^{\prime}$.  The Sloan $B$ band counts were generated by extrapolating between the $u^{\prime}$, $g^{\prime}$, and $r^{\prime}$ bands \citep{2001AJ....122.1104Y}. To facilitate comparisons by future authors we have fit an exponential of the form $N = B 10^{A (magnitude)}$ to our bright end counts where N is in number per square degree per magnitude.  The results of these fits are quoted in table \ref{count slope}.  The number count data are also provided in tables \ref{Raw Number Counts} and \ref{Number Counts}.

\subsection{Selection of High Redshift Objects}

High redshift galaxies have unique colors due to the Lyman break at 912\AA\  and Lyman alpha absorption blueward of 1216\AA.  We can select theses galaxies by choosing three filters which fall blueward of the Lyman break, between the Lyman break and Lyman alpha line, and redward of Lyman alpha respectively.  \citet{1999ApJ...519....1S,1995AJ....110.2519S} successfully uses a set of customized filters to select galaxies at $z \simeq 3$ and $z \simeq 4$.  Their spectroscopic followup is over 80\% success in identifying $z \simeq 3$ galaxies. \citet{2001MNRAS.323..795M} has also used these criteria on the Herschel Deep Field and HDF-N.

The number and type of object selected are strongly dependent on the filters and selection criteria used.  Changing these parameters can severely effect the success in identifying high redshift galaxies and change the redshift range selected.  For comparison to the previous work we need to select the same population as \citet{1999ApJ...519....1S,1995AJ....110.2519S} and \citet{2001MNRAS.323..795M}.  However our band passes differ significantly from theirs.  We corrected this by calculating the expected colors for the galaxies they have selected.  These colors are calculated by integrating the galaxy templates of \citet{1980ApJS...43..393C} and \citet{1996ApJ...467...38K} moderated by our filter response profile.  We correct these templates for Ly$_{\alpha}$ and Ly$_{\beta}$ absorption following the prescription of \citet{1995ApJ...441...18M}.  We also calculate the expected colors of stars using the spectral library of \citet{1998PASP..110..863P}.  We then set a selection criteria which will select similar galaxies to the previous work but avoids stars.  The resulting selection is shown in Figures \ref{UBBRselect} and \ref{BRRIselect} along with color-color tracks for various galaxy types.  The selection is also shown along with our data in figures \ref{UBBR} and \ref{BRRI}.

The surface density we measure using this selection is similar to that measured by other authors \citep{2001MNRAS.323..795M,1999ApJ...519....1S} (see Figures \ref{Udropfig},\ref{Bdropfig} and tables \ref{Udrops},\ref{Bdrops}. Due to our redder $U$ band filter the $z \simeq 3$ selection suffers more contamination from stars and low redshift galaxies than that of \citet{1999ApJ...519....1S,1995AJ....110.2519S}.  We attempted to remove stars by selecting objects with $R < 23.5$ and $R_{fwhm} > 1.3\arcsec$ as describe in \citet{2001A&A...376..756M}, however the contamination from galaxies remains.  At magnitudes fainter than $R = 23$ our measurements are within the one sigma error bars reported by \citet{1999ApJ...519....1S,1995AJ....110.2519S} and \citet{2001MNRAS.323..795M}.  Our B band counts are consistent with but systematically lower than those of \citet{1999ApJ...519....1S} or \citet{2001MNRAS.323..795M} which is likely due to cosmic variance.

The selection at redshifts higher than $z \simeq 4$ is significantly more difficult.  The contamination for $z \simeq 3$ galaxies comes from low redshift galaxies where the co-moving volume is small.  For $z > 5$ galaxies the contamination comes from $z \simeq 1$ where the effective co-moving volume is very large.  Several authors have recently reported the discovery of large numbers of $z > 5$ galaxies using color-color selection \citep{2003PASJ..astroph,2002astroph0212431}.  However we find their two color selection yields a large number of lower redshift galaxies.  Figure \ref{z5selection} shows an example for a redshift 5.5 galaxy.   Elliptical galaxies at $1 < z < 1.5$ have similar colors to a $z > 5.5$ galaxy in the $R$, $I$, and $z^{\prime}$ bands.  The only clear difference is in the bluer or redder bands.  In particular the use of an $(R-I)$ selection will include many low redshift galaxies. A similar problem will occur at redshift 6 using an $(I-Z)$ selection.  There is no easy solution, either one must go more than three magnitudes deeper in the bluest band than the reddest, or obtain deep near IR imaging, both of which are time intensive for large surveys.

\citet{1990ApJ...348..371S} sets out a clear criteria for selecting high redshift galaxies based on the Lyman alpha absorption shown in equation \ref{lbreak}.  

\begin{equation}
	\Delta m_{Ly cont} = 3.8 + 20.3 log _{10} \left( \frac{1+z}{7} \right)
	\label{lbreak}
\end{equation}

For $z > 5$ galaxies it requires $(V-I) > 2.4$ and no detection blueward of the 912\AA\ break.  Using these criteria and our multi-color data we can test the findings of other authors.  To test the level of contamination we have restricted ourselves to $z^{\prime} > 25$ which is bright enough for the expected contaminating sources to be detected at 2$\sigma$ in $U$ or $B$ bands.  

\citet{2003PASJ..astroph} imaged the same area with Suprime-Cam in the same $V$, $I$, and $z^{\prime}$ filter we used.  However they calibrated their data in a different manner.  We adopted $(V-I) \ge 2.4$ and $(V-I) \ge 7(I-z^{\prime})-0.2$ as our selection criteria.  This avoids selecting $z \simeq 1$ galaxies and late type stars.  We find that our counts are consistent with, but lower than the values of \citet{2003PASJ..astroph} (see Figure \ref{Vdropfig} and Tables \ref{Vdrops} and \ref{Vdrops_noB}).  Moving our selection slightly redder in $(I-Z)$ color increases the number of objects selected, making our numbers more consistent with \citet{2003PASJ..astroph}.  However all the additional objects are detected in $U$ or $B$ band.  Even when objects detected in $U$ or $B$ are removed, many of the remaining objects in the ACS GOODS field morphologically appear to be stars (this will be quantified elsewhere).  These results suggest that a combination of $z \simeq 1$ galaxies and late type stars are contaminating the $z > 5.5$ selection.

Figure \ref{RIIZ} shows that the selection criteria of \citet{2002astroph0212431} is very likely selecting low redshift galaxies or stars.  We find that 95\% of the objects selecting using this criteria are detected at 2$\sigma$ in $V$.  This selection is particularly problematic since it may include star forming galaxies at $z < 2$ where [\ion{O}{3}] or [\ion{O}{2}] can be mistaken for Lyman $\alpha$.  Finally the $(I-Z) > 1.5$ selection used by \citet{2003astroph0302212}, selects objects which are detected at 2$\sigma$ in $V$ band 53\% of the time.  In contrast only 14\% of objects selecting using the $(B-R)$, $(R-I)$ selection were detected in $U$ band.  Based on our number counts the probability of an object at 28$^{th}$ magnitude falling into a 3\asec aperture by chance is 37\% in $V$ band and 4\% in $U$ band. This implies that color selection may be overestimating the number of high redshift galaxies and hence the star formation rates at $z > 5$.  We shall give a more extensive discussion of this issue with photometric redshift estimates in a future paper \citep{Capak2003}.

\subsection{Conclusion}
We have compiled a deep, multi-color catalog over 0.2 square degrees in the HDF-N region. These data will provide an invaluable basis for understanding the formation and evolution of galaxies by providing a large sample of galaxies which can be studied in various ways.  The raw number counts and counts of $z \simeq 3$, and $z \simeq 4$ galaxies agree with those of other authors.  However the selection of $z > 5$ galaxies suffers from significant contamination by low redshift galaxies.  The current deep blue imaging is required to reliably selected these galaxies and hence estimate the star formation rate.

\acknowledgements
We'd like to thank Nick Kaiser for providing his software and supporting it. We'd also like to thank H. McCracken and N. Metcalfe for providing their number count data.  The work of C. McNally and D. Stern was carried out at Jet Propulsion Laboratory, California Institute of Technology, under a contract with NASA. This work was supported by NASA grant DF1-2001X(L.L.C) and NSF grants AST00-71208 (E.M.H), AST 00-84816 (L.L.C.), and AST 99-83783 (A.J.B.).

\clearpage

\clearpage
\begin{figure}
\begin{center}
\begin{tabular}{ccc}
\includegraphics[scale=0.4]{capak.fig2-left.ps}& &\includegraphics[scale=0.4]{capak.fig2-right.ps}\\
\end{tabular}
\end{center}
\caption{Astrometric residuals with respect to the radio catalog of \citet{2000ApJ...533..611R}.  The inner box on the left hand side marks the central 0.2 square degrees in the residual map. The vectors point in the direction of the residual and have and amplitude of 60 times the residual in arc seconds. \label{astro-radiofig}}
\end{figure}

\clearpage

\clearpage

\clearpage

\clearpage

\clearpage

\addtocounter{figure}{1}

\clearpage
\begin{figure}
\epsscale{1}
\plotone{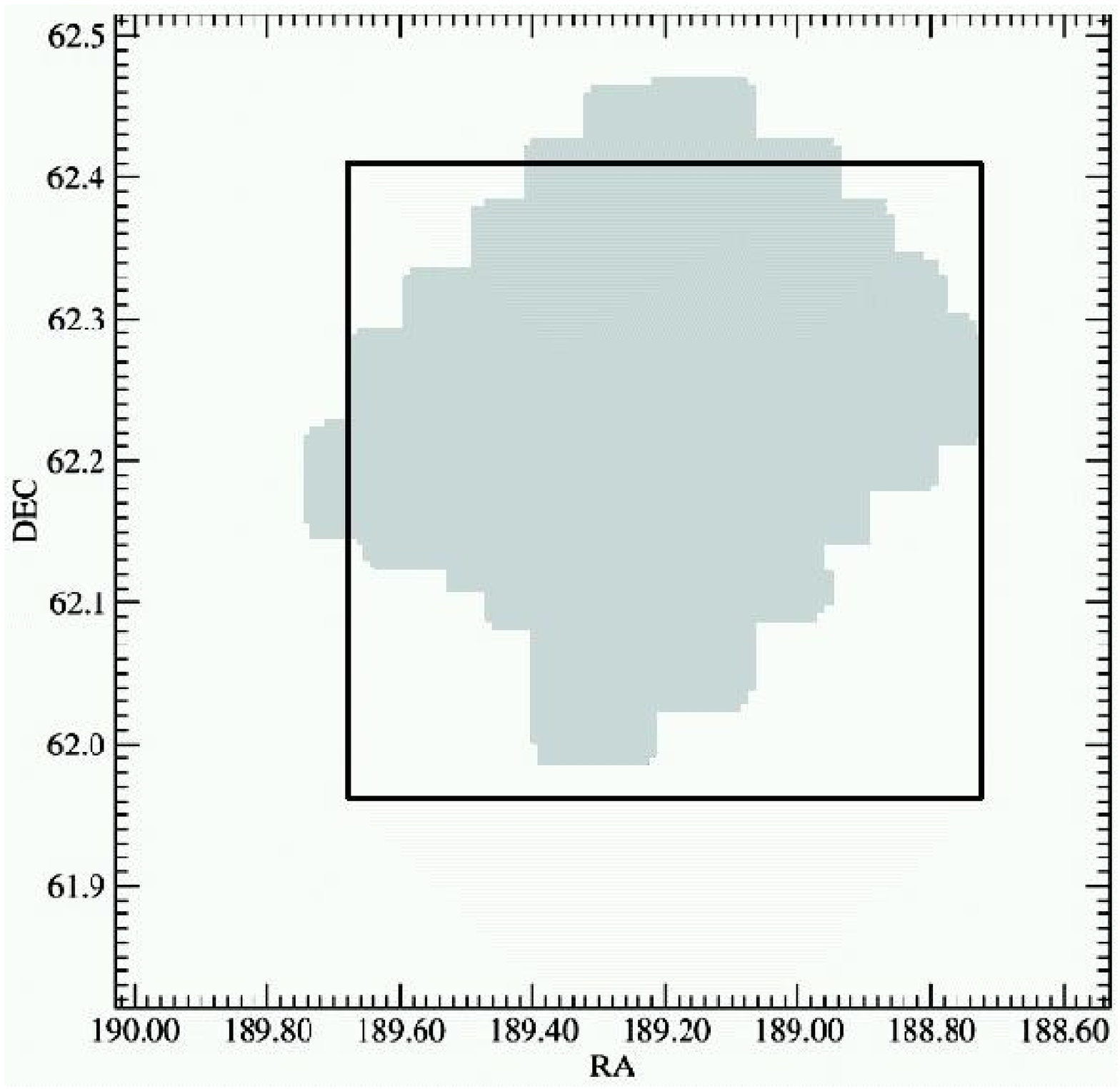}
\caption{The area with $HK^{\prime}$ coverage is shaded in grey, the catalog area is shown by the black outline, please note that the $HK^{\prime}$ data does not cover the whole catalog area.}
\label{HKarea}
\end{figure}

\clearpage
\begin{figure}
\epsscale{1}
\plotone{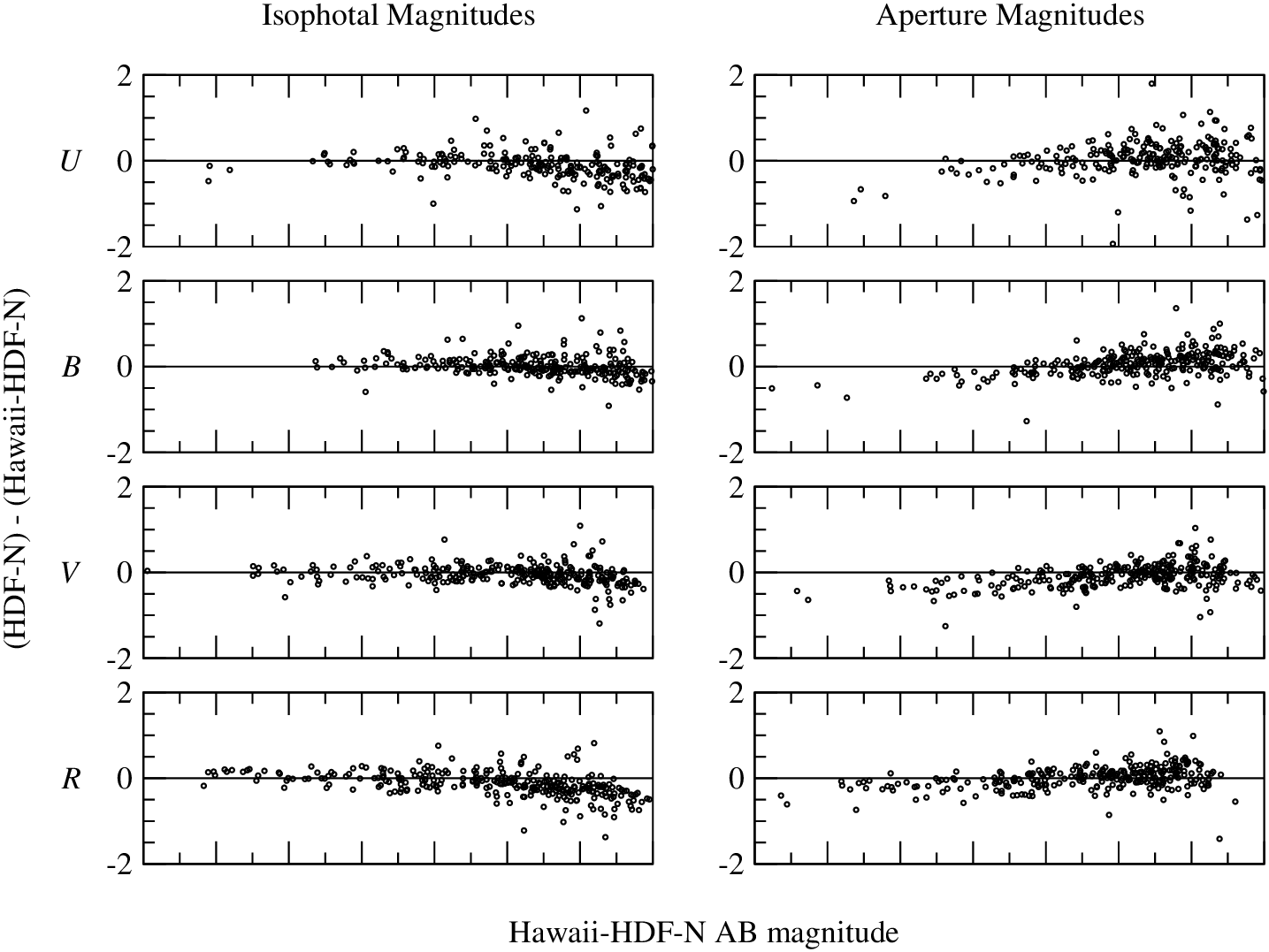}
\caption{Comparison of our photometry to that of \citet{1999ApJ...513...34F} from the HDF-N.  The magnitudes have been converted to our filter system by linearly interpolating the flux between the HDF-N filters. The RMS scatter between our isophotal magnitudes and those of Fern\'{a}ndez-Soto et al. is 0.19 magnitudes on average which corresponds to an average error of 0.03 magnitudes in the zero points.\label{abs phot1}}
\end{figure}

\clearpage
\begin{figure}
\epsscale{1}
\plotone{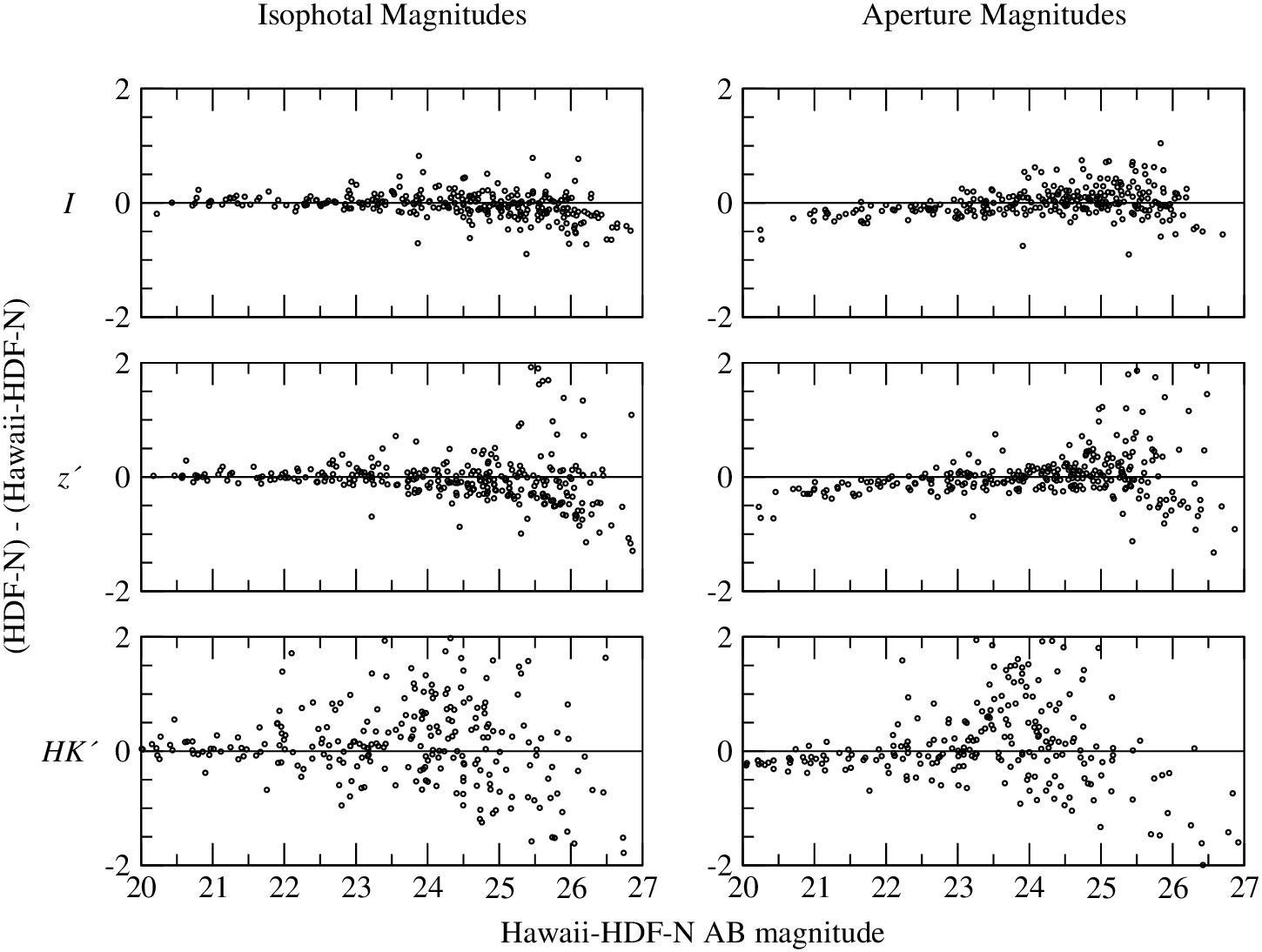}
\caption{Comparison of our photometry to that of \citet{1999ApJ...513...34F} from the HDF-N.  The magnitudes have been converted to our filter system by linearly interpolating the flux between the HDF-N filters. The RMS scatter between our isophotal magnitudes and those of Fern\'{a}ndez-Soto et al. is 0.19 magnitudes on average which corresponds to an average error of 0.03 magnitudes in the zero points.\label{abs phot2}}
\end{figure}

\clearpage
\begin{figure}
\plotone{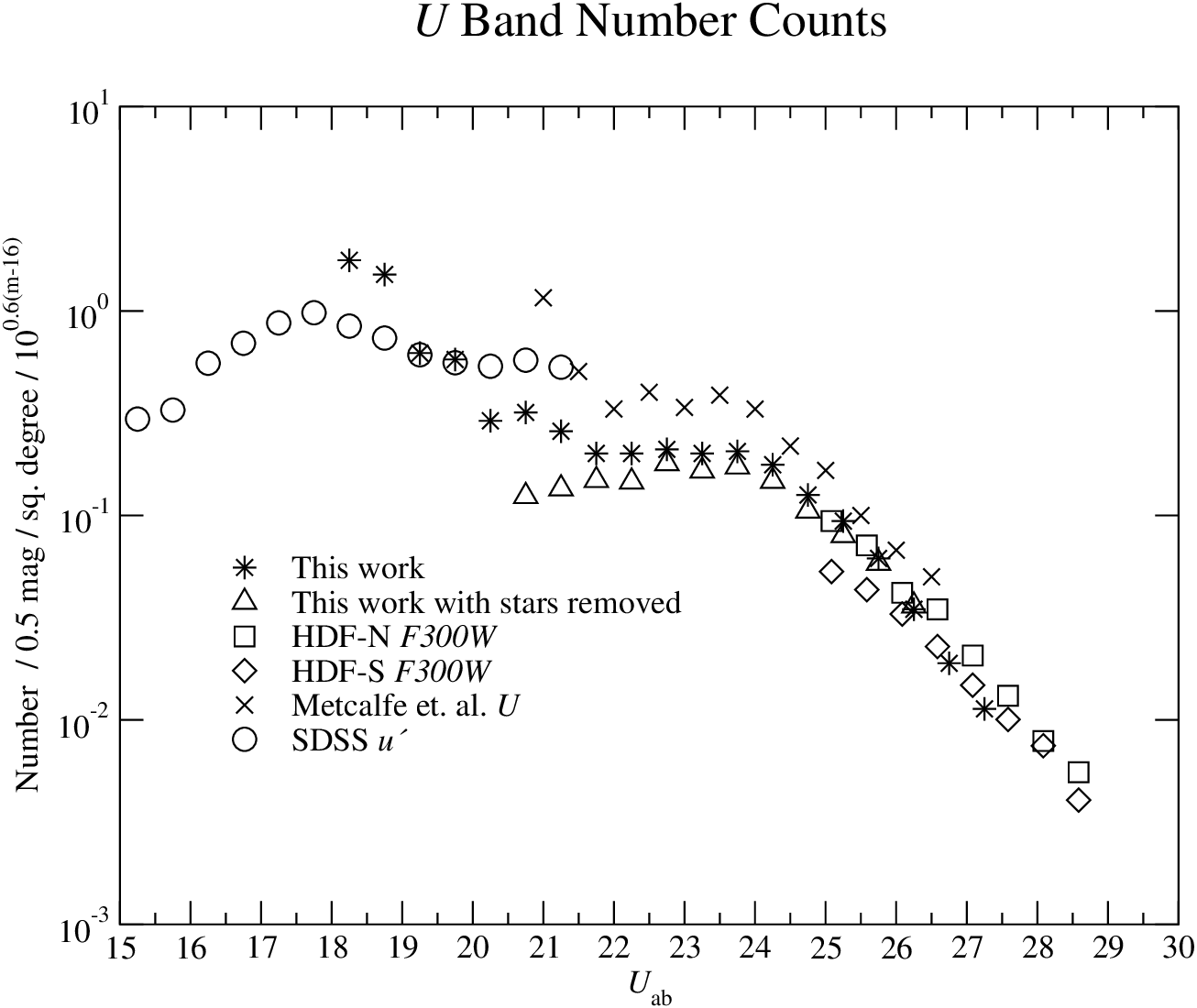}
\caption{U band number counts from this work, the Herschel Deep Field \citep{2001MNRAS.323..795M}, the HDF-N \citep{2001MNRAS.323..795M}, and the Sloan Digital Sky Survey (SDSS) \citep{2001AJ....122.1104Y}. Objects were identified as stars if they had a value of 0.9 or greater in the Sextractor star galaxy-separator.\label{Ucountfig}}
\end{figure}

\clearpage
\begin{figure}
\plotone{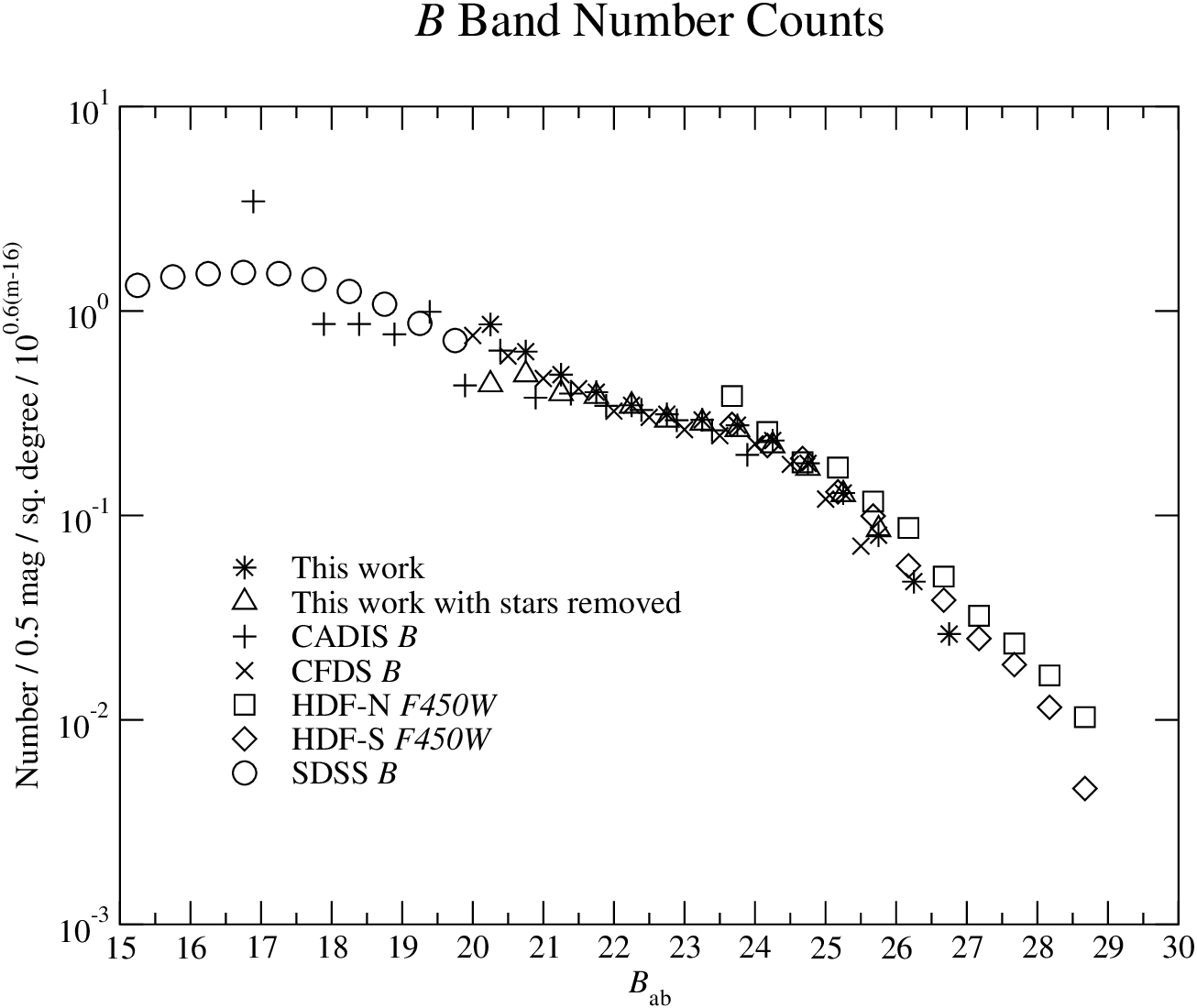}
\caption{B band number counts from this work, the Calar Alto Deep Imaging Survey \citep{2001A&A...368..787H}, the Canada France Deep Survey (CFDS)\citep{2001A&A...376..756M}, the HDF-N \citep{2001MNRAS.323..795M}, and the Sloan Digital Sky Survey (SDSS) \citep{2001AJ....122.1104Y}. Objects were identified as stars if they had a value of 0.9 or greater in the Sextractor star galaxy-separator.\label{Bcountfig}}
\end{figure}

\clearpage
\begin{figure}
\plotone{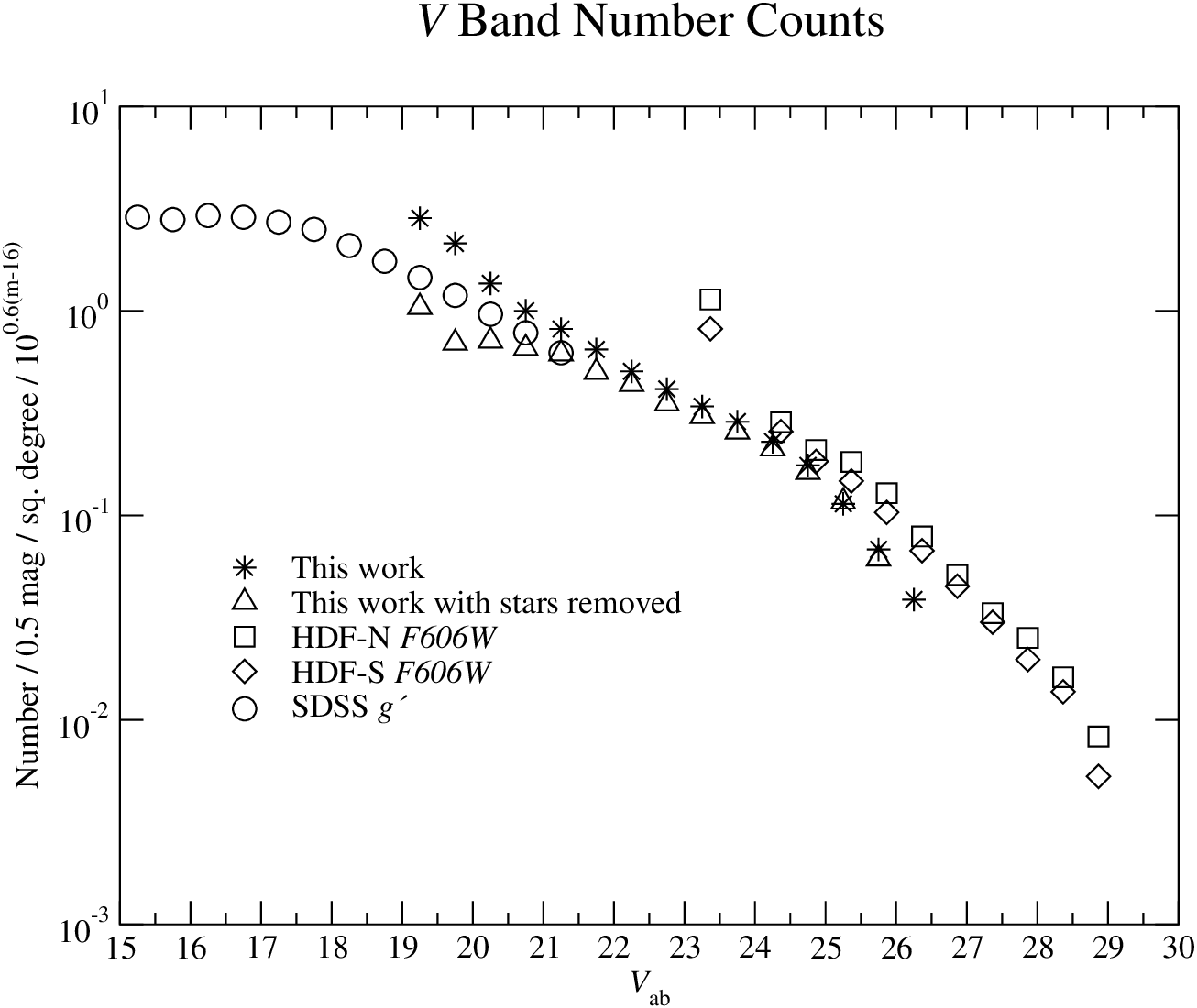}
\caption{V band number counts from this work, the HDF-N \citep{2001MNRAS.323..795M}, and the Sloan Digital Sky Survey (SDSS) \citep{2001AJ....122.1104Y}.\label{Vcountfig}}
\end{figure}

\clearpage
\begin{figure}
\plotone{capak.fig10.ps}
\caption{R band number counts from this work, the Calar Alto Deep Imaging Survey \citep{2001A&A...368..787H}, the Herschel Deep Field \citep{2001MNRAS.323..795M}, the HDF-N \citep{2001MNRAS.323..795M}, and the Sloan Digital Sky Survey (SDSS) \citep{2001AJ....122.1104Y}. Objects were identified as stars if they had a value of 0.9 or greater in the Sextractor star galaxy-separator.\label{Rcountfig}}
\end{figure}

\clearpage
\begin{figure}
\plotone{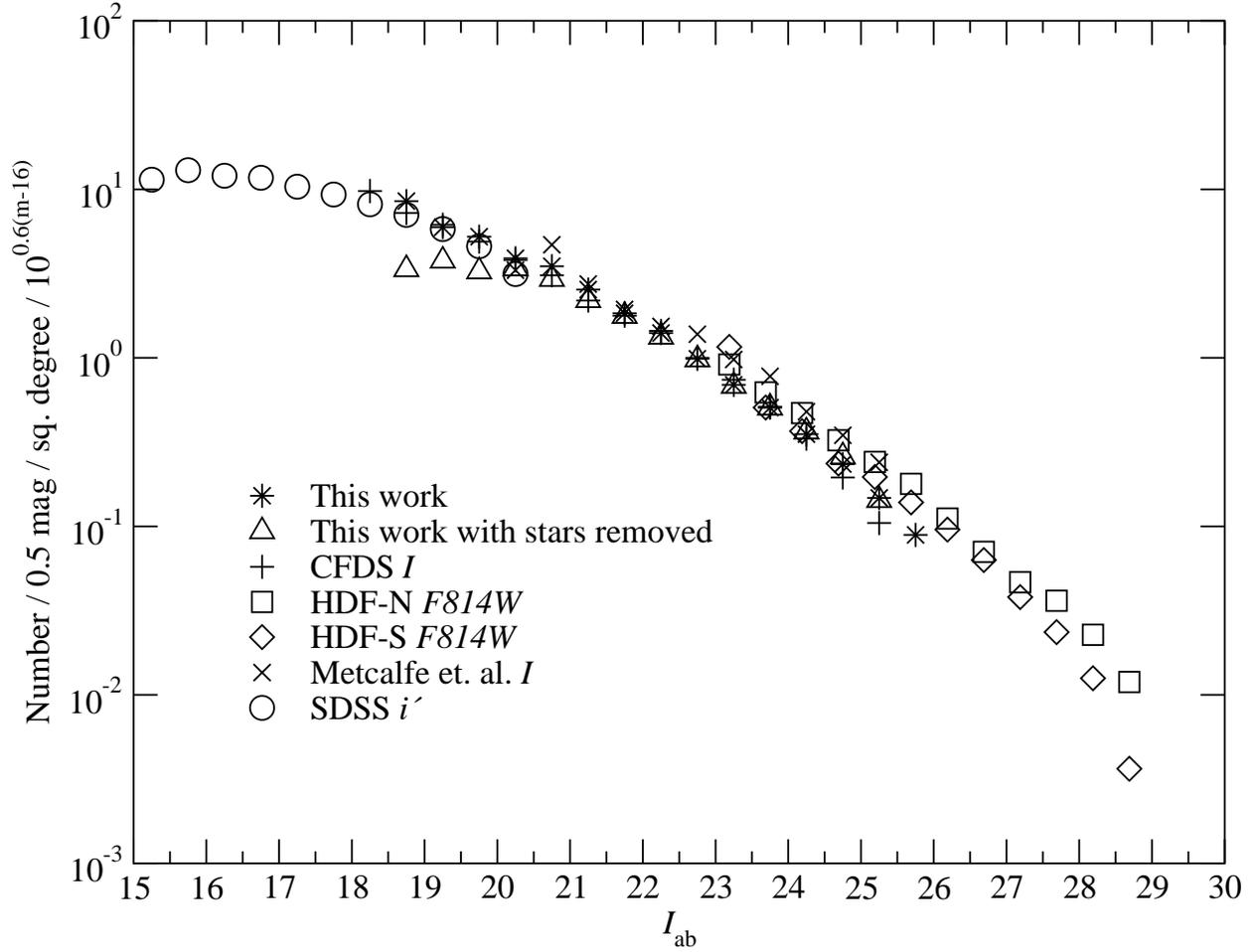}
\caption{I band number counts from this work, the Canada France Deep Survey (CFDS)\citep{2001A&A...376..756M}, the Herschel Deep Field \citep{2001MNRAS.323..795M}, the HDF-N \citep{2001MNRAS.323..795M}, and the Sloan Digital Sky Survey (SDSS) \citep{2001AJ....122.1104Y}. Objects were identified as stars if they had a value of 0.9 or greater in the Sextractor star galaxy-separator.\label{Icountfig}}
\end{figure}

\clearpage
\begin{figure}
\plotone{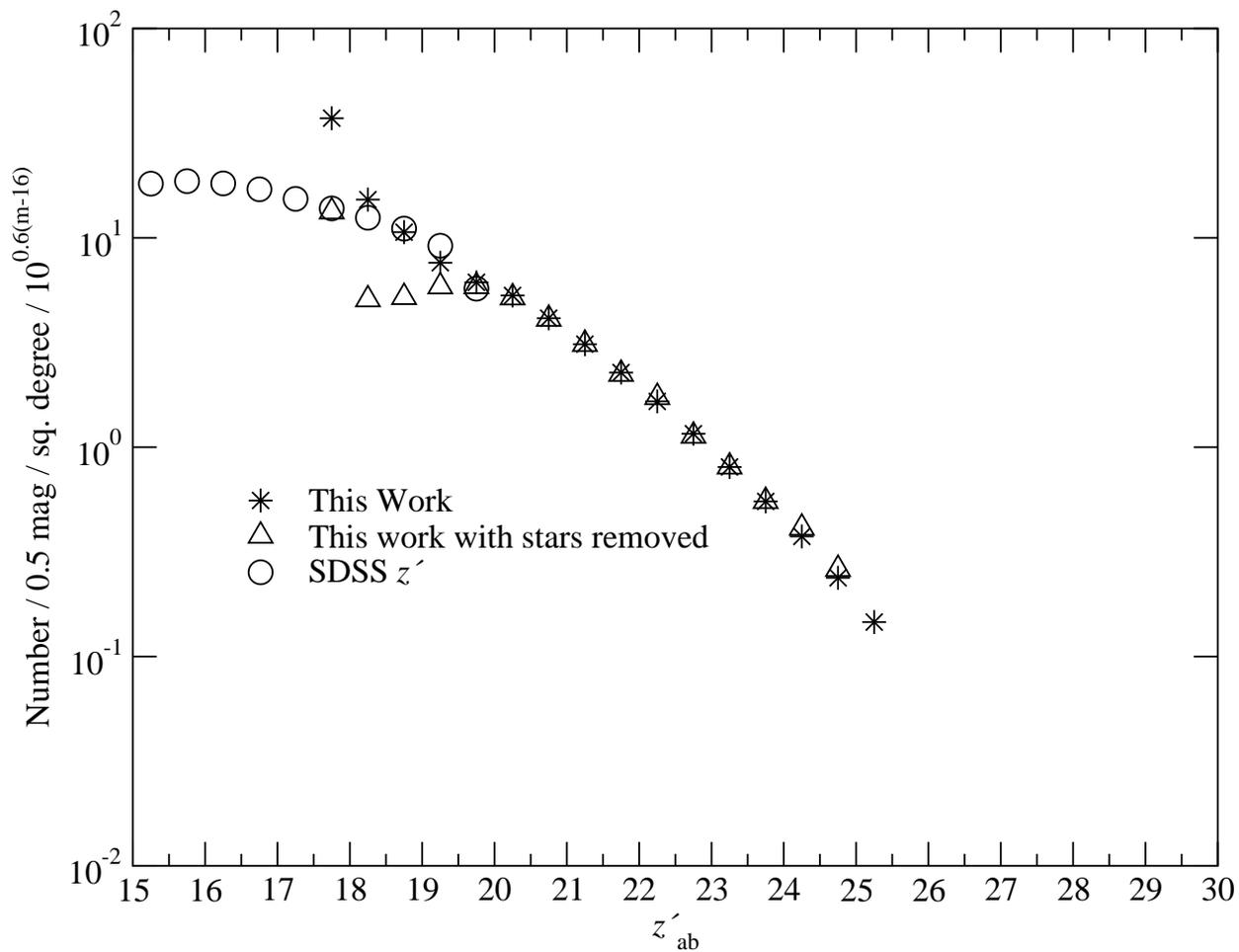}
\caption{$z^{\prime}$ band number counts from this work and the Sloan Digital Sky Survey (SDSS) \citep{2001AJ....122.1104Y}. Objects were identified as stars if they had a value of 0.9 or greater in the Sextractor star galaxy-separator.\label{Zcountfig}}
\end{figure}

\clearpage
\begin{figure}
\plotone{capak.fig13.ps}
\caption{K band number counts from this work, the Chandra Deep Field South (CDF-S) \citep{2001A&A...375....1S}, the HDF-N \citep{2001A&A...375....1S}, and several other authors \citep{1998ApJ...505...50B, 1997ApJ...475..445M, 1995ApJ...438L..13D,1994ApJ...420L...1S,1993ApJ...415L...9G} are over plotted.  An offset of 1.859 was applied to Vega magnitudes to convert them to the AB system. \label{Kcountfig}}
\end{figure}

\clearpage
\begin{figure}
\plotone{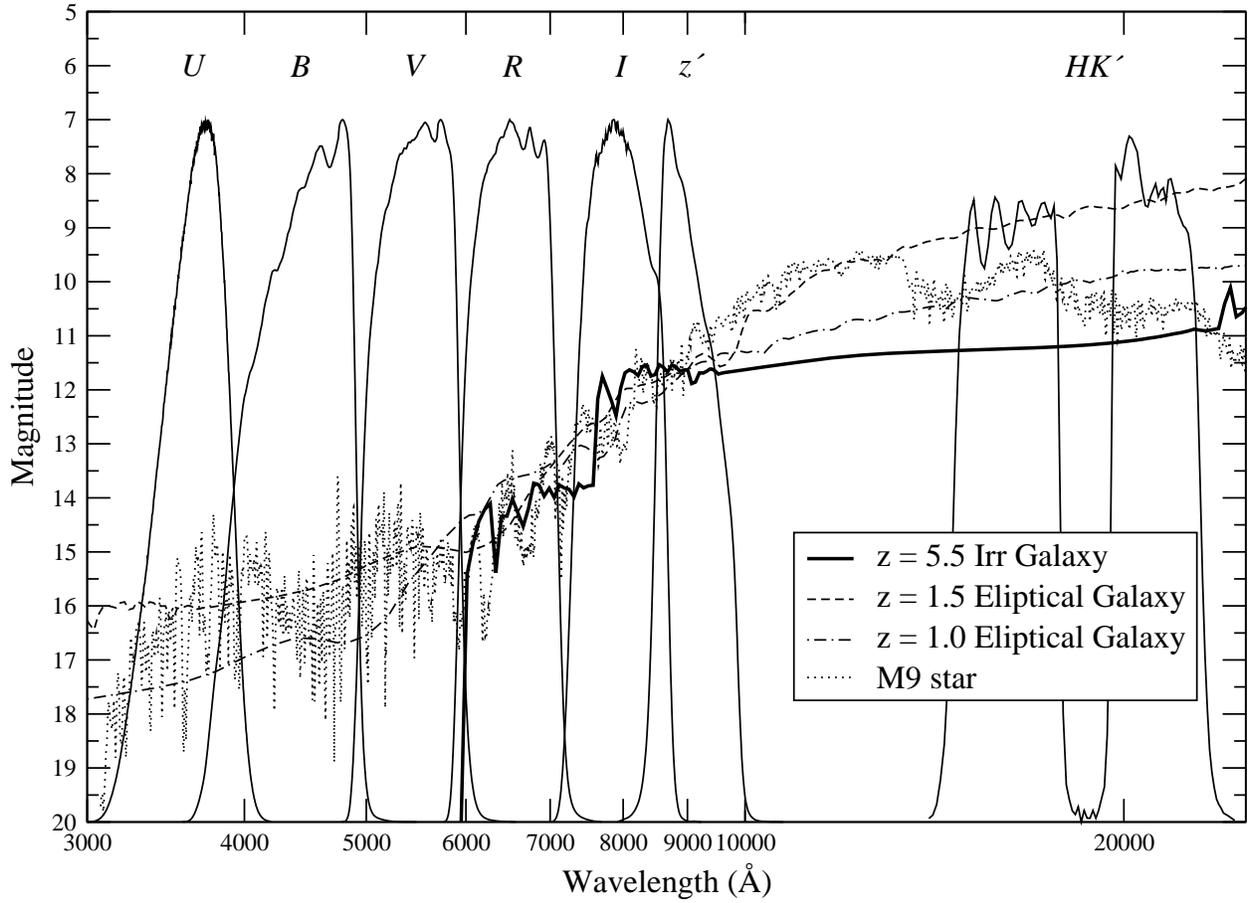}
\caption{The expected Spectral Energy Distribution (SED) of a $ z =$ 5.5 star forming, $z =$ 1.5 Elliptical, and $z =$ 1.0 Elliptical galaxy from the \citet{1980ApJS...43..393C} library are shown along with an M9 star from \citet{1998PASP..110..863P}.  These SED's were corrected for intergalactic absorption using \citet{1995ApJ...441...18M}. The filter profiles of our survey are overlaid for comparison.  Note the similarity of the SED's in $R$, $I$, and $z^{\prime}$ bands.  Note that differentiating the $z = 5.5$ Irr from the $z = 1.0$ or $z = 1.5$ elliptical requires multi-color information and is most sensitive in the $U$, $B$, and near-IR bands.  The maximum information comes from the largest wavelength range sample. \label{z5selection}}
\end{figure}

\clearpage
\begin{figure}
\includegraphics[scale=0.7]{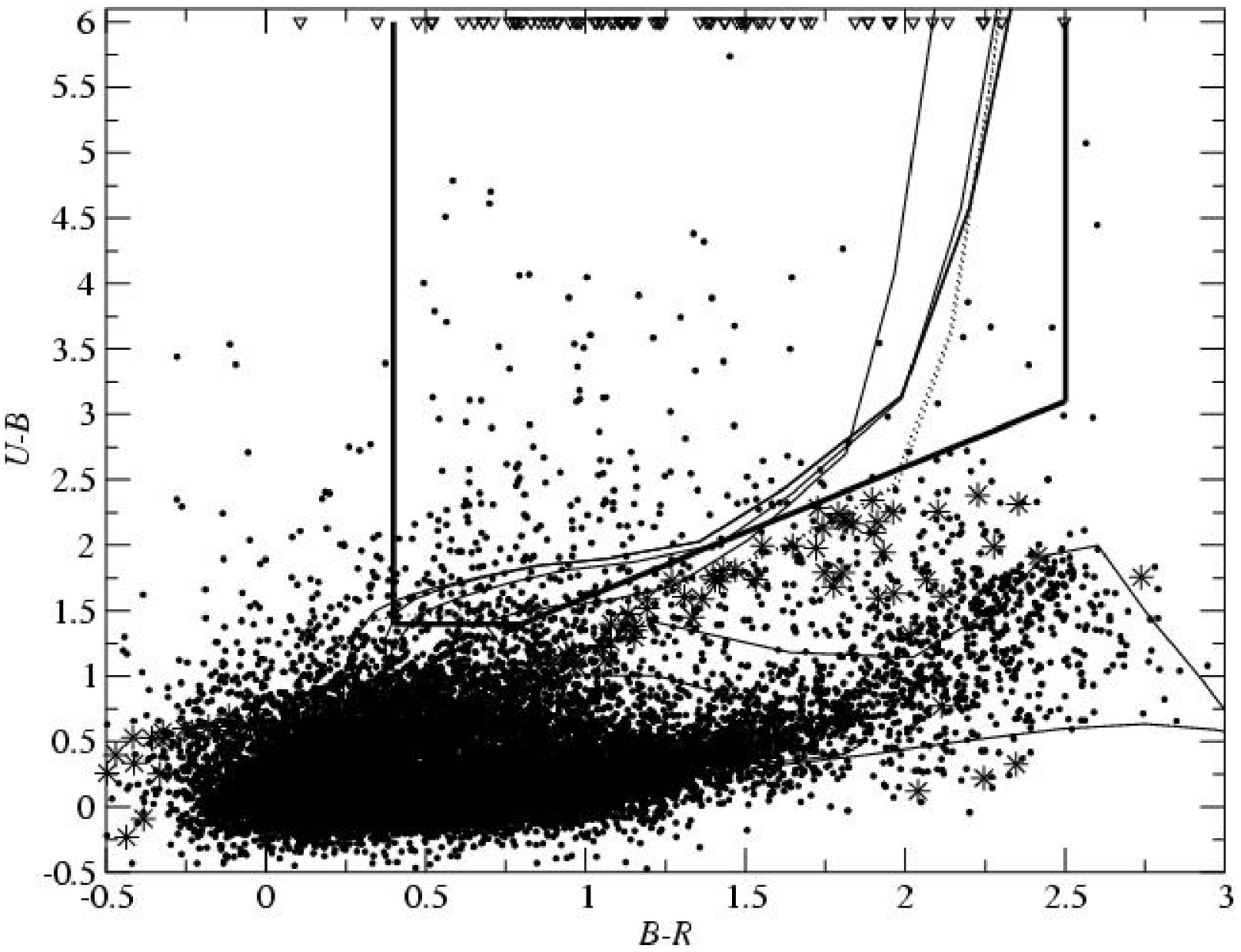}
\caption{$(B-R)$ vs. $(U-B)$ color-color plot for all objects with $B_{AB} > 26$. The region indicated by heavy lines in the top right is where we expect to find $z \simeq 3$ objects similar to those of \citet{1995AJ....110.2519S}. The light solid lines show the expected evolution of \citet{1980ApJS...43..393C} galaxy templates and the dotted lines the \citet{1996ApJ...467...38K} SB2 and SB3 galaxy templates. The expected colors of stars from \citet{1998PASP..110..863P} are over plotted as stars.  The downward facing triangles at $(U-B) = 6$ are objects which were not detected in U.\label{UBBR}}
\end{figure}

\clearpage
\begin{figure}
\includegraphics[scale=0.7]{capak.fig16.ps}
\caption{$(B-R)$ vs. $(U-B)$ color-color plot with galaxy tracks. The region indicated by heavy lines in the top right is where we expect to find $z \simeq 3$ objects similar to those of \citet{1995AJ....110.2519S}. The light solid lines show the expected evolution of \citet{1980ApJS...43..393C} galaxy templates and the dotted lines show the \citet{1996ApJ...467...38K} SB2 and SB3 galaxy templates. The expected colors of stars from \citet{1998PASP..110..863P} are over plotted as stars.  Circles indicate galaxies in the $2.5 < z < 3.5$ redshift range selected by \citet{1995AJ....110.2519S}. \label{UBBRselect}}
\end{figure}

\clearpage
\begin{figure}
\includegraphics[scale=0.7]{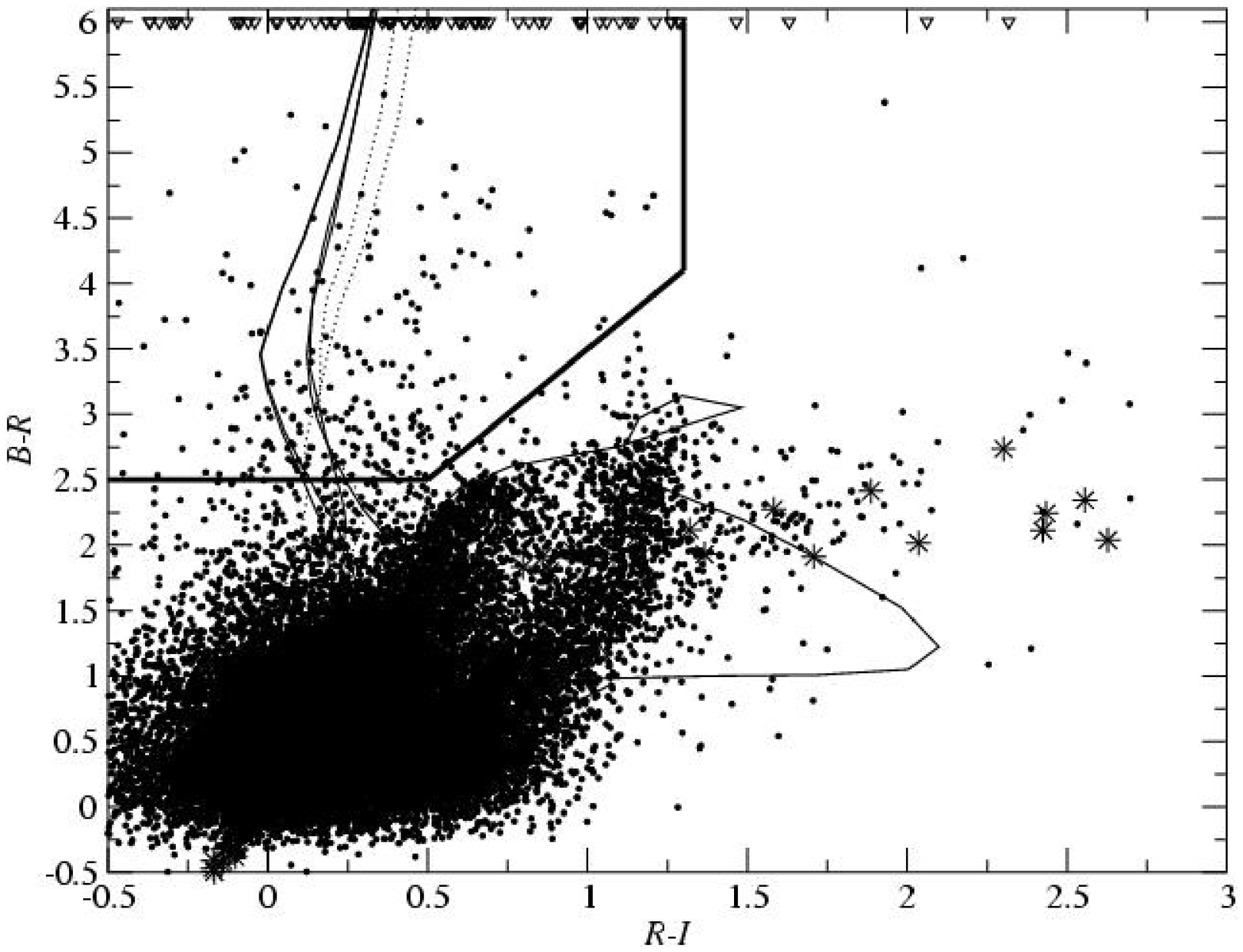}
\caption{$(R-I)$ vs. $(B-R)$ color-color plot for all objects with $R_{AB} > 26$. The region indicated by heavy lines in the top left is where we expect to find $z \simeq 4$ objects similar to those of \citet{1999ApJ...519....1S}. The light solid lines show the expected evolution of \citet{1980ApJS...43..393C} galaxy templates and the dotted lines show the \citet{1996ApJ...467...38K} SB2 and SB3 galaxy templates. The expected colors of stars from \citet{1998PASP..110..863P} are over plotted as stars.  The downward facing triangles  at $(B-R) = 6$ are objects which were not detected in B. \label{BRRI}}
\end{figure}

\clearpage
\begin{figure}
\includegraphics[scale=0.7]{capak.fig18.ps}
\caption{$(R-I)$ vs. $(B-R)$ color-color plot with galaxy tracks. The region indicated by heavy lines in the top left is where we expect to find $z \simeq 4$ objects similar to those of \citet{1999ApJ...519....1S}. The light solid lines show the expected evolution of \citet{1980ApJS...43..393C} galaxy templates and the dotted lines show the \citet{1996ApJ...467...38K} SB2 and SB3 galaxy templates. The expected colors of stars from \citet{1998PASP..110..863P} are over plotted as stars.  Circles indicate galaxies in the $3.5 < z < 4.8$ redshift range selected by \citet{1995AJ....110.2519S}.\label{BRRIselect}}
\end{figure}

\clearpage
\begin{figure}
\includegraphics[scale=0.7]{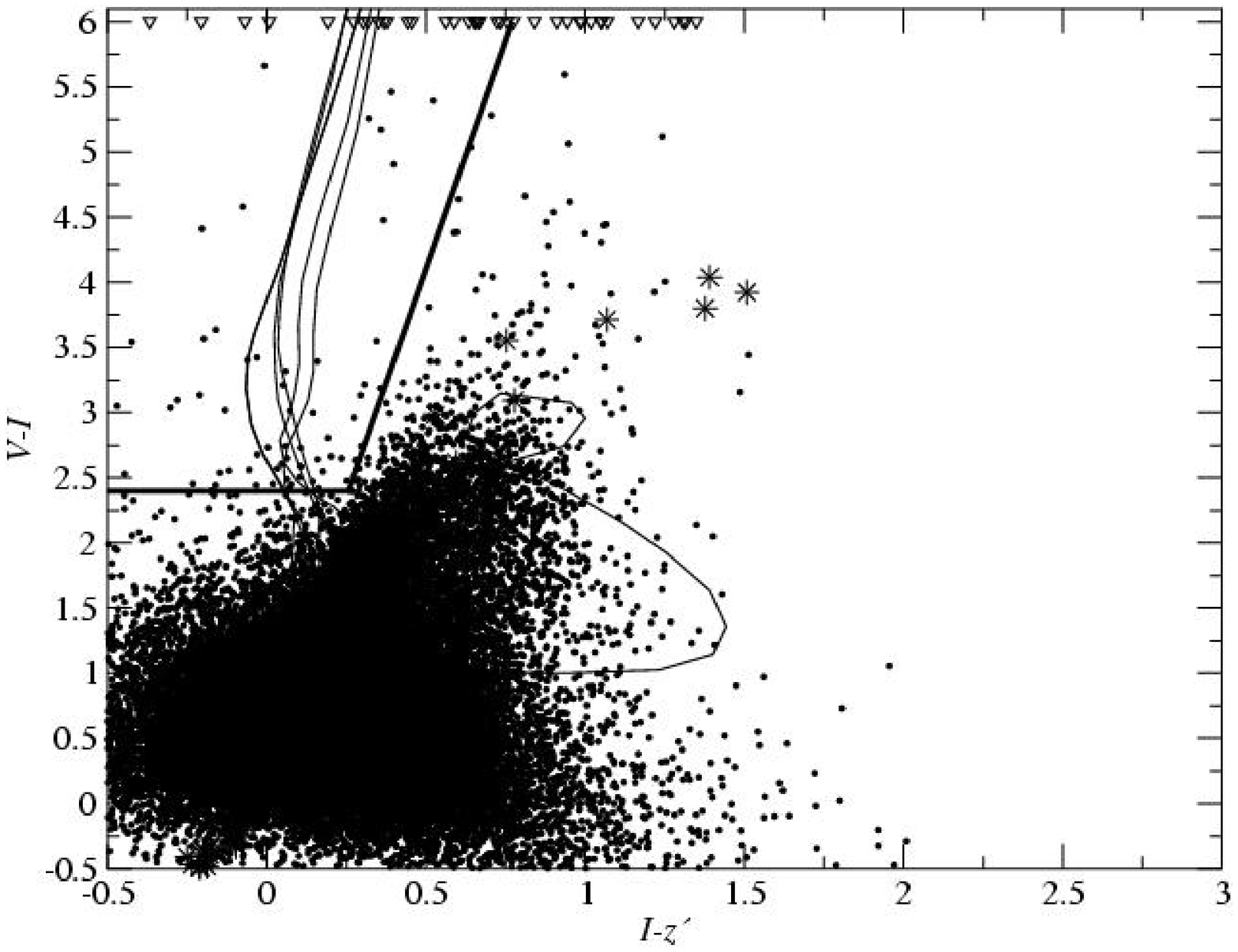}
\caption{$(I-z^{\prime})$ vs. $(V-I)$ color-color plot for all detected objects. The region indicated by heavy lines in the top left is where we expect to find $z \simeq 5.5$ objects based on the criteria of \citet{2003PASJ..astroph}. We have moved the cuts to $(V-I)>=2.4$ and $(V-I) > 7(I-Z)-0.2$ to avoid contamination from lower redshift galaxies and late type stars as well as account for differences in the photometric system. The expected colors of stars from \citet{1998PASP..110..863P} are over plotted as stars. The downward facing triangles  at $(V-I) = 6$ are objects which were not detected in V.\label{VIIZ}}
\end{figure}

\clearpage
\begin{figure}
\includegraphics[scale=0.7]{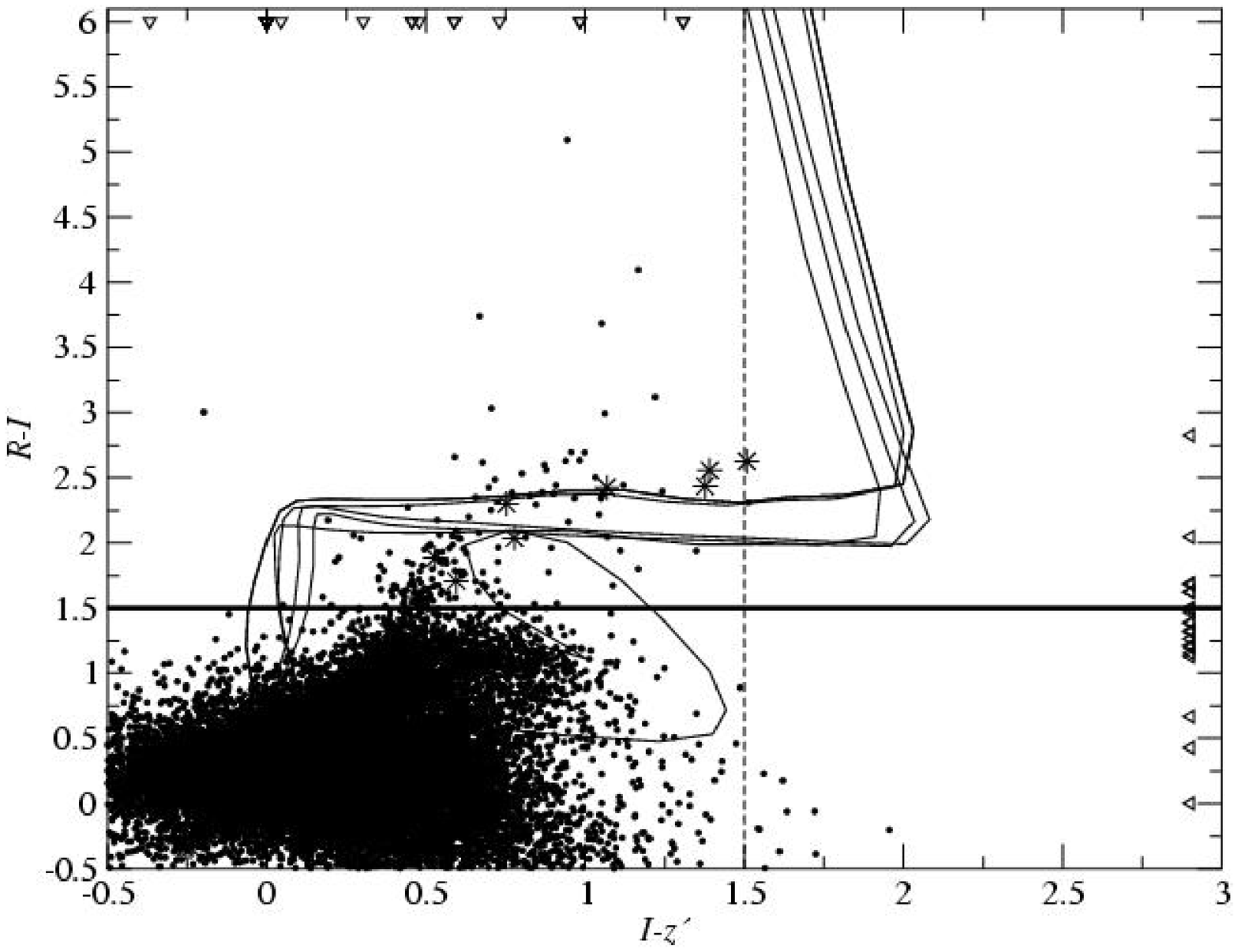}
\caption{$(I-z^{\prime})$  vs. $(R-I)$ color-color plot for all detected objects. The heavy line at $(R-I) = 1.5$ indicates the selection criteria of \citet{2002astroph0212431} and the dashed line at $(I-Z) = 1.5$ is the selection criteria of \citet{2003astroph0302212}.  The light solid lines show the expected evolution of \citet{1980ApJS...43..393C} galaxy templates and the \citet{1996ApJ...467...38K} SB2 and SB3 galaxy templates. The expected colors of stars from \citet{1998PASP..110..863P} are over plotted as stars. The downward pointing triangles  at $(R-I) = 6$ are objects which were not detected in R and leftward pointing triangles at $(I-Z) = 6$ are objects which were not detected in I.\label{RIIZ}}
\end{figure}

\clearpage
\begin{figure}
\plotone{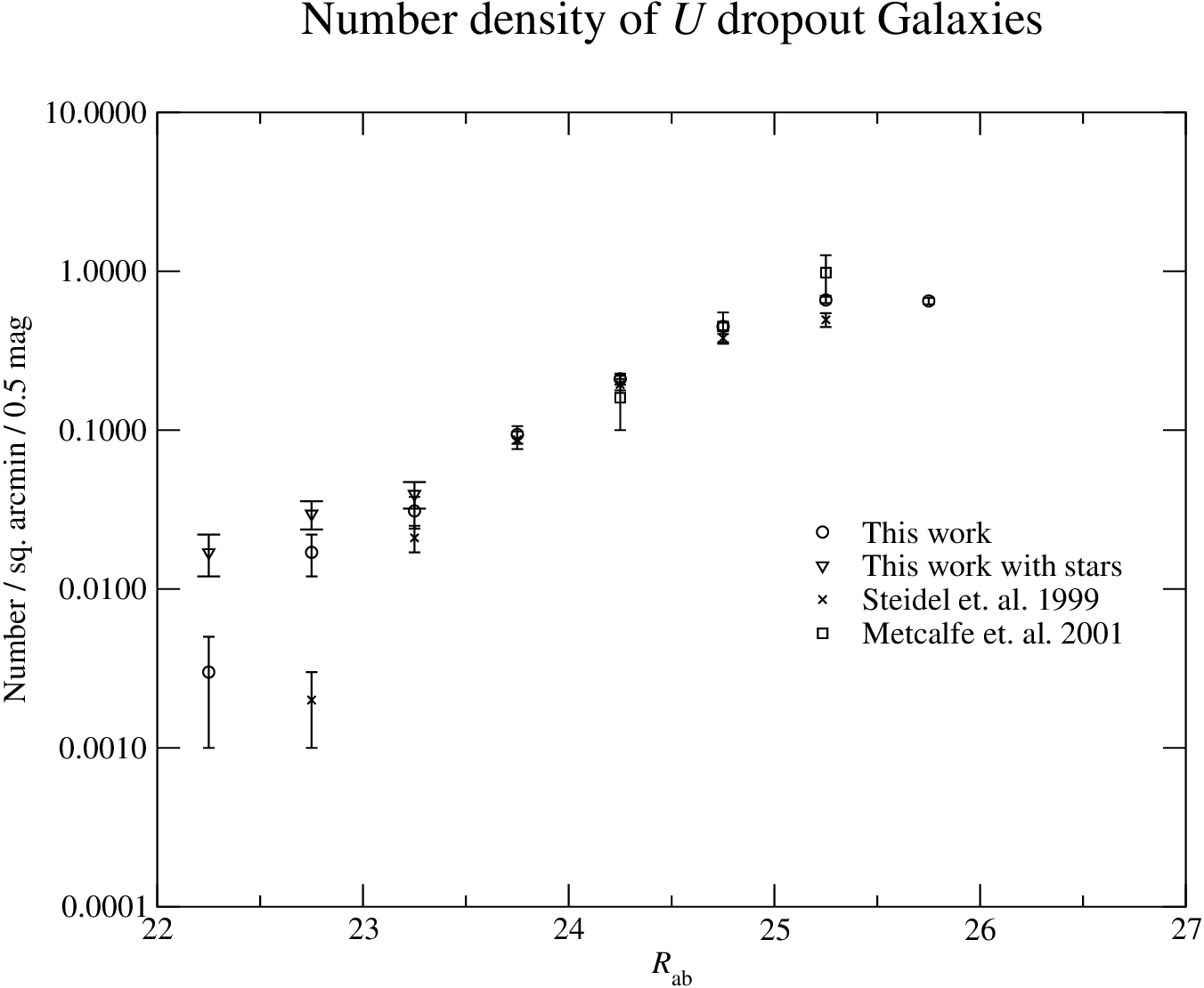}
\caption{The raw sky density of $U$ band dropouts uncorrected for contamination or incompleteness are shown.  Error bars are 1$\sigma$ poison fluctuations for \citet{2001MNRAS.323..795M} and our data.  The errors for \citet{1999ApJ...519....1S} include an estimate of cosmic variance from their multiple fields.  Much of the variation between the counts is likely due to cosmic variance. \label{Udropfig}}
\end{figure}

\clearpage
\begin{figure}
\plotone{capak.fig22.ps}
\caption{The raw sky density of $B$ band dropouts uncorrected for contamination or incompleteness are shown. Error bars are 1$\sigma$ poison fluctuations for \citet{2001MNRAS.323..795M} and our data.  The errors for \citet{1999ApJ...519....1S} include an estimate of cosmic variance from their multiple fields. Much of the variation between the counts is likely due to cosmic variance. \label{Bdropfig}}
\end{figure}

\clearpage
\begin{figure}
\plotone{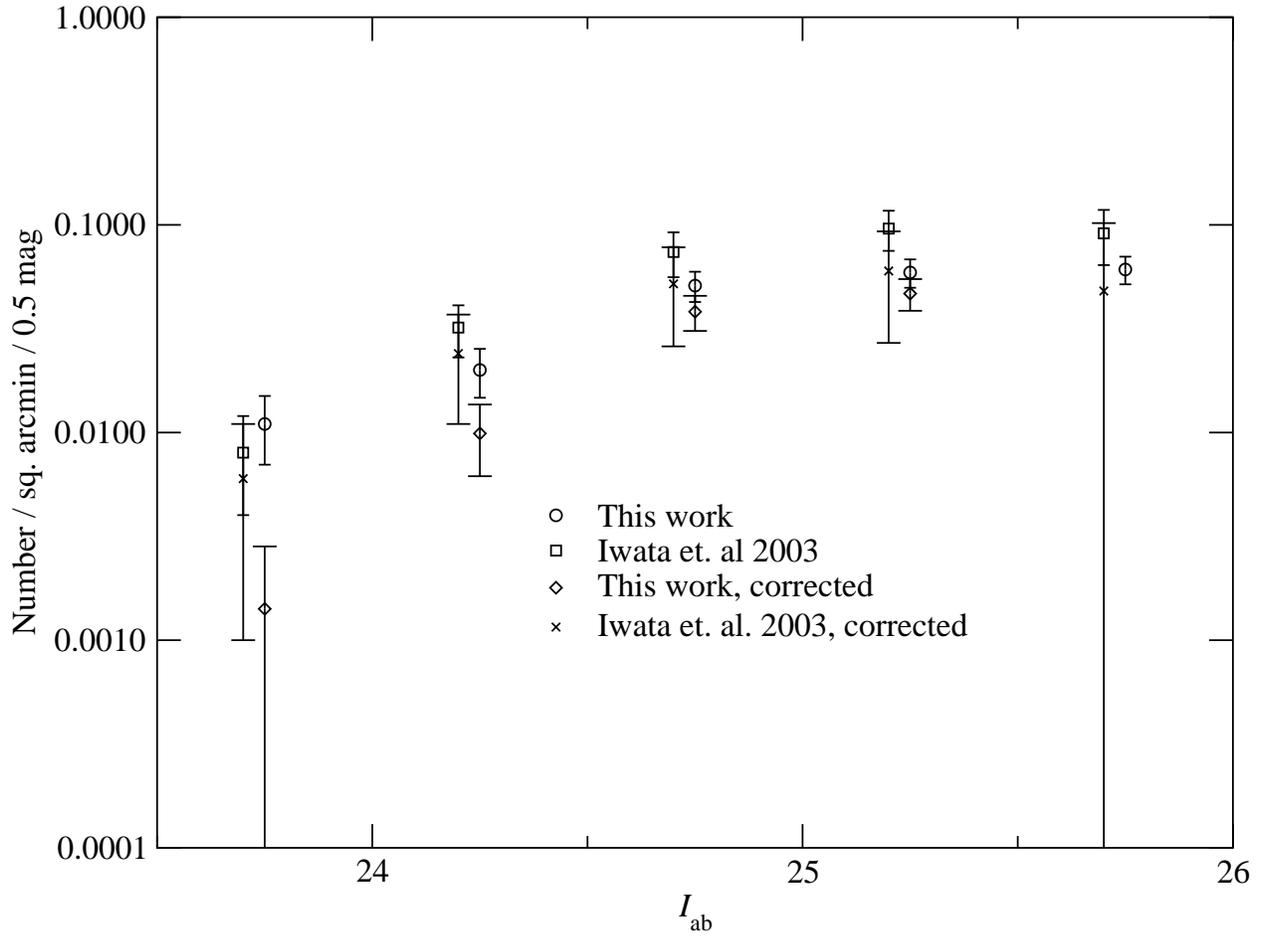}
\caption{The raw sky density of $V$ band dropouts and an estimate of the density corrected for contamination are shown for the Hawaii-HDF-N. Neither measurement is corrected for incompleteness.  Error bars are 1$\sigma$ poison fluctuations.  The counts of \citet{2003PASJ..astroph} are higher due to a higher rate of contamination. \label{Vdropfig}}
\end{figure}

\clearpage
\begin{deluxetable}{ccccl}
\scriptsize
\rotate
\tablecaption{Telescopes used, Exposure Times, and Dates of Observations\label{data details1}}
\tablehead{
\colhead{Band} &   
\colhead{Integration\tablenotemark{a}}&
\colhead{Telescope} &
\colhead{Typical}&
\colhead{Date of}\\
\colhead{} &   
\colhead{Time (h)}&
\colhead{Used} &
\colhead{Integration (s)}&
\colhead{Observations}}
\startdata 
$U$   & 28.5 & KPNO 4m     & 1800 & 2002, March, 9-13\\
$B$   & 1.7  & Subaru 8.3m & 600  & 2001, February, 27,28\\
$V$   & 6.4  & Subaru 8.3m & 1200 & 2001, February, 23,24\\
      & 2.0  &             & 600  & 2001, March 21-23\\
$R$   & 5.2  & Subaru 8.3m & 480  & 2001, February, 27,28\\
      &      &             & 600  & 2001, March, 21-23\\
$I$   & 2.9  & Subaru 8.3m & 300  & 2001, February, 23,24\\
      &      &             & 300  & 2001, March, 21-23\\
      &      &             & 300  & 2002, April, 5,11-14\\
$z^{\prime}$  & 3.9  & Subaru 8.3m & 180  & 2001, February, 23,24,27,28\\
      &      &             & 240  & 2000, March, 21-23\\
$HK^{\prime}$ & 0.43 & UH 2.2m     & 120  & 1999-2002\\
\enddata
\tablenotetext{a}{This is the average integration time for a pixel. This varies significantly in the $HK^{\prime}$ image.}
\end{deluxetable}

\clearpage
\begin{deluxetable}{cccccl}
\scriptsize
\rotate
\tablecaption{Data Quality, Depth, and Coverage\label{data details2}}
\tablehead{
\colhead{Band} & 
\colhead{Seeing} & 
\colhead{$5\sigma$ limit\tablenotemark{a}} &  
\colhead{Total area}&
\colhead{Deep area}&
\colhead{Date of}\\
\colhead{} & 
\colhead{arcsecond} & 
\colhead{(AB mag)} &  
\colhead{Covered (deg$^{2}$)}&
\colhead{Covered (deg$^{2}$)}&
\colhead{Observations}}
\startdata 
$U$   & 1.26 & 27.1 & 0.40 & 0.36 & 2002, March, 9-13\\
$B$   & 0.71 & 26.9 & 0.27 & 0.20 & 2001, February, 27,28\\
$V$   & 0.71 & 26.8 & 0.27 & 0.20 & 2001, February, 23,24\\
      & 1.18 &      & 0.40 & 0.20 & 2001, March 21-23\\
$R$   & 1.11 & 26.6 & 0.27 & 0.20 & 2001, February, 27,28\\
      &      &      & 0.40 & 0.20 & 2001, March, 21-23\\
$I$   & 0.72 & 25.6 & 0.27 & 0.20 & 2001, February, 23,24\\
      &      &      & 0.40 & 0.20 & 2001, March, 21-23\\
      &      &      & 0.40 & 0.20 & 2002, April, 5,11-14\\
$z^{\prime}$  & 0.67 & 25.4 & 0.27 & 0.20 & 2001, February, 23,24,27,28\\
      &      &      & 0.40 & 0.20 & 2000, March, 21-23\\
$HK^{\prime}$ & 0.87 & 22.1 \tablenotemark{b}& 0.11 & 0.11 & 1999-2002\\
\enddata
\tablenotetext{a}{Measured by randomly placing $10^{4}$ 3\asec diameter apertures on an image with detected objects masked out.  The apertures were placed at least 3\asec away from the positions of detected object. }
\tablenotetext{b}{This is the mode of the depth across the field.  In a 9\amin x 9\amin area around the HDF-N the $5\sigma$ limit is 22.8.}
\end{deluxetable}

\clearpage
\begin{deluxetable}{ccccccc}
\scriptsize
\rotate
\tablecaption{Photometric Information\label{Photometry Offsets}}
\tablehead{
\colhead{Band} & 
\colhead{Zeropoint} &
\colhead{Offsets from} &
\colhead{Aperture}&
\colhead{Offsets to} &
\colhead{Central} &
\colhead{Bandpass \AA}\\
\colhead{ } & 
\colhead{Error\tablenotemark{a}} &
\colhead{SED fitting} &
\colhead{Correction \tablenotemark{b}} &
\colhead{Vega system\tablenotemark{c}} &  
\colhead{Wavelength \AA} &
\colhead{}}
\startdata 
$U$           & 0.031 & -0.008 & -0.311 &  0.719 & 3647.65  & 387.16\\
$B$           & 0.028 &  0.000 & -0.159 & -0.077 & 4427.60  & 622.05\\
$V$           & 0.028 & -0.005 & -0.282 &  0.023 & 5471.22  & 577.36\\
$R$           & 0.026 &  0.061 & -0.262 &  0.228 & 6534.16  & 676.39\\
$I$           & 0.016 &  0.000 & -0.126 &  0.453 & 7975.89  & 792.89\\
$z^{\prime}$  & 0.018 & -0.065 & -0.147 &  0.532 & 9069.21  & 802.63\\
$HK^{\prime}$ & 0.042 & -0.181 & -0.233 &  1.595 & 18947.38 & 5406.22\\
\enddata
\tablenotetext{a}{Includes the estimated error of 0.01 magnitudes in the \citet{1999ApJ...513...34F} catalog.}
\tablenotetext{b}{Used in the calculation of number counts.}
\tablenotetext{c}{Calculated by integrating an A0 star SED multiplied by our filter profile.}
\end{deluxetable}

\clearpage
\begin{deluxetable}{cl}
\tabletypesize{\scriptsize}
\tablecaption{Contents of Shape File\label{shape cat}}
\tablehead{
\colhead{Column Number} & 
\colhead{Shape Catalog Item}}
\startdata 
1  & ID                  \\
2  & RA(J2000)           \\
3  & DEC(J2000)          \\
4  & x position on image \\
5  & y position on image \\
6  & $U$ FWHM \tablenotemark{a}             \\
7  & $U$ Semi-Major Axis \tablenotemark{a}  \\
8  & $U$ Semi-Minor Axis \tablenotemark{a}  \\
9  & $B$ FWHM \tablenotemark{a}             \\
10 & $B$ Semi-Major Axis \tablenotemark{a}  \\
11 & $B$ Semi-Minor Axis \tablenotemark{a}  \\
12 & $V$ FWHM \tablenotemark{a}             \\
13 & $V$ Semi-Major Axis \tablenotemark{a}  \\
14 & $V$ Semi-Minor Axis \tablenotemark{a}  \\
15 & $R$ FWHM \tablenotemark{a}             \\
16 & $R$ Semi-Major Axis \tablenotemark{a}  \\
17 & $R$ Semi-Minor Axis \tablenotemark{a}  \\
18 & $I$ FWHM \tablenotemark{a}             \\
19 & $I$ Semi-Major Axis \tablenotemark{a}  \\
20 & $I$ Semi-Minor Axis \tablenotemark{a} \\
21 & $z^{\prime}$ FWHM \tablenotemark{a}             \\
22 & $z^{\prime}$ Semi-Major Axis \tablenotemark{a}  \\
23 & $z^{\prime}$ Semi-Minor Axis \tablenotemark{a}  \\
24 & $HK^{\prime}$ FWHM \tablenotemark{a}           \\
25 & $HK^{\prime}$ Semi-Major Axis\tablenotemark{a} \\
26 & $HK^{\prime}$ Semi-Minor Axis\tablenotemark{a} \\
\enddata
\tablenotetext{a}{This is the output from the Sextractor FWHM-IMAGE measurement. The FWHM is calculated in a different manor than the Semi-Major and Semi-Minor axis which which are outputs from the A-IMAGE and B-IMAGE measurements. All three of these measurements were made on the un-smoothed images with respect to centroids in the detection images and have been scaled to arcseconds.  A value of 0 indicates that the measurement could not be determined.  For more information on these parameters see \citet{1996A&AS..117..393B}.}
\end{deluxetable}

\clearpage
\begin{deluxetable}{cll}
\tabletypesize{\scriptsize}
\tablecaption{Contents of Flag File\label{flag cat}}
\tablehead{
\colhead{Column Number} & 
\colhead{Flag Catalog Item} &
\colhead{Comments}}
\startdata 
1   & ID             &            \\
2   & Bad            & Set to 1 if the FWHM is 0 in the detecting band\\
3   & $U$ Saturated    & Set to 1 if any pixel is above saturation  the limit\\
4   & $B$ Saturated\tablenotemark{a}& Set to 1 if $B < 20.49$\\
5   & $V$ Saturated\tablenotemark{a}& Set to 1 if $V < 19.85$\\
6   & $R$ Saturated\tablenotemark{a}& Set to 1 if $R < 20.25$ \\
7   & $I$ Saturated\tablenotemark{a}& Set to 1 if $I < 20.15$\\
8   & $z^{\prime}$ Saturated\tablenotemark{a}& Set to 1 if $Z < 19.45$\\
9   & $HK^{\prime}$ Saturated  & Set to 1 if any pixel is above saturation  the limit\\
10  & N Overlapping  & Number of objects detected within 3\asec of this object\\
\enddata
\tablenotetext{a}{Saturated pixels behave in a very non-linear fashion on Suprime-Cam and can drop in value after reaching the saturation limit.  We adopted these magnitude cuts by measuring the magnitude at which point sources became non-Gaussian.}
\end{deluxetable}

\clearpage
\begin{deluxetable}{cl}
\tabletypesize{\scriptsize}
\tablecaption{Contents of Magnitude File\label{mag cat}}
\tablehead{
\colhead{Column Number} & 
\colhead{Magnitude Catalog Item\tablenotemark{a}}}
\startdata 
1   & ID                      \\
2   & $U$ Aperture Magnitude    \\
3   & $U$ Aperture Magnitude Error      \\
4   & $U$ Isophotal Magnitude   \\
5   & $U$ Isophotal Magnitude Error       \\
6   & $B$ Aperture Magnitude    \\
7   & $B$ Aperture Magnitude Error      \\
8   & $B$ Isophotal Magnitude   \\
9   & $B$ Isophotal Magnitude Error       \\
10  & $V$ Aperture Magnitude    \\
11  & $V$ Aperture Magnitude Error      \\
12  & $V$ Isophotal Magnitude   \\
13  & $V$ Isophotal Magnitude Error       \\
14  & $R$ Aperture Magnitude    \\
15  & $R$ Aperture Magnitude Error      \\
16  & $R$ Isophotal Magnitude   \\
17  & $R$ Isophotal Magnitude Error       \\
18  & $I$ Aperture Magnitude    \\
19  & $I$ Aperture Magnitude Error      \\
20  & $I$ Isophotal Magnitude   \\
21  & $I$ Isophotal Magnitude Error       \\
22  & $z^{\prime}$ Aperture Magnitude    \\
23  & $z^{\prime}$ Aperture Magnitude Error      \\
24  & $z^{\prime}$ Isophotal Magnitude   \\
25  & $z^{\prime}$ Isophotal Magnitude Error       \\
26  & $HK^{\prime}$ Aperture Magnitude  \\
27  & $HK^{\prime}$ Aperture Magnitude Error    \\
28  & $HK^{\prime}$ Isophotal Magnitude \\
29  & $HK^{\prime}$ Isophotal Magnitude Error     \\
\enddata
\end{deluxetable}

\clearpage
\begin{deluxetable}{cl}
\tabletypesize{\scriptsize}
\tablecaption{Contents of Flux File\label{flux cat}}
\tablehead{
\colhead{Column Number} & 
\colhead{Flux Catalog Item}}
\startdata 
1   & ID\\            
2   & $U$ Aperture Flux\\         
3   & $U$ Aperture Flux Error\\   
4   & $U$ Aperture Background\\
5   & $U$ Isophotal Flux\\    
6   & $U$ Isophotal Flux Error\\
7   & $U$ Isophotal Background\\
8   & $B$ Aperture Flux\\ 
9   & $B$ Aperture Flux Error\\
10  & $B$ Aperture Background\\
11  & $B$ Isophotal Flux\\  
12  & $B$ Isophotal Flux Error\\
13  & $B$ Isophotal Background\\ 
14  & $V$ Aperture Flux\\        
15  & $V$ Aperture Flux Error\\  
16  & $V$ Aperture Background\\   
17  & $V$ Isophotal Flux\\
18  & $V$ Isophotal Flux Error\\  
19  & $V$ Isophotal Background\\
20  & $R$ Aperture Flux\\       
21  & $R$ Aperture Flux Error\\  
22  & $R$ Aperture Background\\
23  & $R$ Isophotal Flux\\
24  & $R$ Isophotal Flux Error\\  
25  & $R$ Isophotal Background\\ 
26  & $I$ Aperture Flux\\ 
27  & $I$ Aperture Flux Error\\
28  & $I$ Aperture Background\\
29  & $I$ Isophotal Flux\\
30  & $I$ Isophotal Flux Error\\
31  & $I$ Isophotal Background\\
32  & $z^{\prime}$ Aperture Flux\\
33  & $z^{\prime}$ Aperture Flux Error\\
34  & $z^{\prime}$ Aperture Background\\
35  & $z^{\prime}$ Isophotal Flux\\
36  & $z^{\prime}$ Isophotal Flux Error\\
37  & $z^{\prime}$ Isophotal Background\\
38  & $HK^{\prime}$ Aperture Flux\\
39  & $HK^{\prime}$ Aperture Flux Error\\
40  & $HK^{\prime}$ Aperture Background\\
41  & $HK^{\prime}$ Isophotal Flux\\
42  & $HK^{\prime}$ Isophotal Flux Error\\
43  & $HK^{\prime}$ Isophotal Background\\
\enddata
\tablenotetext{a}{All flux values are in nano-Janskys (nJy).}
\end{deluxetable}

\clearpage
\begin{deluxetable}{cccccc}
\tablecaption{Fit of the form  $N = B 10^{A (AB~magnitude)}$ to the number counts\tablenotemark{a} \label{count slope}}
\tablehead{
\colhead{Band} &
\colhead{Fit Range in AB magnitudes}&
\colhead{A} &
\colhead{Error in A} &
\colhead{$log_{10}(B)$} &
\colhead{Error in $log_{10}(B)$}}
\startdata 
 $U$           & 20.0-24.5 & 0.526 & 0.017 & -8.61 & 0.37\\
 $B$           & 20.0-25.5 & 0.450 & 0.008 & -6.67 & 0.18\\
 $V$           & 20.0-25.0 & 0.402 & 0.004 & -5.50 & 0.09\\
 $R$           & 20.0-25.0 & 0.361 & 0.004 & -4.36 & 0.08\\
 $I$           & 20.0-25.0 & 0.331 & 0.008 & -3.54 & 0.17\\
 $z^{\prime}$  & 20.0-25.0 & 0.309 & 0.006 & -2.98 & 0.13\\
 $HK^{\prime}$ & 16.0-19.0 & 0.378 & 0.037 & -4.15 & 0.65\\
 $HK^{\prime}$ & 19.0-22.0 & 0.393 & 0.019 & -4.31 & 0.39\\
\enddata
\tablenotetext{a}{N has units of number degree$^{-2}$ 0.5 mag$^{-1}$}
\end{deluxetable}

\clearpage
\begin{deluxetable}{cccccccc}
\tablecaption{Raw number counts in N degree$^{-2}$ 0.5 mag$^{-1}$\label{Raw Number Counts}}
\tabletypesize{\scriptsize}
\rotate
\tablehead{
\colhead{Magnitude} & 
\colhead{$U$} & 
\colhead{$B$} &  
\colhead{$V$} &
\colhead{$R$} &
\colhead{$I$} &
\colhead{$z^{\prime}$} &
\colhead{$HK^{\prime}$}}
\startdata 
 18.25	&	41 $\pm$ 14	&	\nodata	        &	\nodata	        &	\nodata		&	\nodata		&	\nodata		&	536 $\pm$ 70 \\ 
 18.75	&	46 $\pm$ 15	&	\nodata	        &	\nodata	        &	\nodata		&	\nodata		&	454 $\pm$ 48	&	869 $\pm$ 89 \\ 
 19.25	&	77 $\pm$ 20	&	\nodata	        &	\nodata	        &	\nodata		&	511 $\pm$ 51	&	656 $\pm$ 58	&	1673 $\pm$ 124 \\ 
 19.75	&	98 $\pm$ 22	&	\nodata	        &	346 $\pm$ 42	&	\nodata		&	914 $\pm$ 68	&	1012 $\pm$ 72	&	2996 $\pm$ 166 \\ 
 20.25	&	103 $\pm$ 23	&	315 $\pm$ 40	&	459 $\pm$ 48	&	852 $\pm$ 66	&	1229 $\pm$ 79	&	1787 $\pm$ 96	&	4632 $\pm$ 206 \\ 
 20.75	&	263 $\pm$ 36	&	485 $\pm$ 50	&	723 $\pm$ 61	&	1265 $\pm$ 80	&	2190 $\pm$ 106	&	2800 $\pm$ 120	&	7148 $\pm$ 257 \\ 
 21.25	&	371 $\pm$ 43	&	712 $\pm$ 60	&	1100 $\pm$ 75	&	2190 $\pm$ 106	&	3259 $\pm$ 129	&	3797 $\pm$ 140	&	12317 $\pm$ 337 \\ 
 21.75	&	557 $\pm$ 53	&	1224 $\pm$ 79	&	1777 $\pm$ 95	&	3048 $\pm$ 125	&	4918 $\pm$ 159	&	6044 $\pm$ 176	&	15655 $\pm$ 380 \\ 
 22.25	&	1048 $\pm$ 73	&	2071 $\pm$ 103	&	2588 $\pm$ 115	&	4551 $\pm$ 153	&	7113 $\pm$ 191	&	8854 $\pm$ 213	&	\\ 
 22.75	&	2371 $\pm$ 110	&	3435 $\pm$ 133	&	4282 $\pm$ 148	&	6850 $\pm$ 188	&	10301 $\pm$ 230	&	12135 $\pm$ 250	&	\\ 
 23.25	&	4096 $\pm$ 145	&	6468 $\pm$ 182	&	7067 $\pm$ 191	&	10585 $\pm$ 233	&	13984 $\pm$ 268	&	16624 $\pm$ 293	&	\\ 
 23.75	&	8389 $\pm$ 208	&	11887 $\pm$ 247	&	11484 $\pm$ 243	&	16268 $\pm$ 289	&	20380 $\pm$ 324	&	22622 $\pm$ 341	&	\\ 
 24.25	&	14284 $\pm$ 271	&	19647 $\pm$ 318	&	18727 $\pm$ 311	&	24224 $\pm$ 353	&	29488 $\pm$ 390	&	33802 $\pm$ 417	& \\ 
 24.75	&	19982 $\pm$ 321	&	30361 $\pm$ 396	&	28858 $\pm$ 386	&	35925 $\pm$ 430	&	42827 $\pm$ 470	&	45235 $\pm$ 483	&	\\ 
 25.25	&	29824 $\pm$ 392	&	45064 $\pm$ 482	&	40869 $\pm$ 459	&	48474 $\pm$ 500	&	51016 $\pm$ 513	&	44718 $\pm$ 480	&\\ 
 25.75	&	42135 $\pm$ 466	&	60542 $\pm$ 559	&	44073 $\pm$ 477	&	61410 $\pm$ 563	&			&		&	\\ 
 26.25	&	51093 $\pm$ 513	&	70141 $\pm$ 601	&	27918 $\pm$ 379	&	51863 $\pm$ 517	&			&		&	\\ 
 26.75	&	39361 $\pm$ 450	&	55831 $\pm$ 537	&			&			&			&		&	\\ 
 27.25	&	21568 $\pm$ 333	& 			& 			&			& 			&			& \\ 
\enddata
\end{deluxetable}

\clearpage      
\begin{deluxetable}{ccccccc}
\tablecaption{Number counts with stars removed in N degree$^{-2}$ 0.5 mag$^{-1}$\label{Number Counts}}
\tabletypesize{\scriptsize}
\rotate
\tablehead{
\colhead{Magnitude} & 
\colhead{$U$} & 
\colhead{$B$} &  
\colhead{$V$} &
\colhead{$R$} &
\colhead{$I$} &
\colhead{$z^{\prime}$}}
\startdata 
 18.25	&	5 $\pm$ 5	&	\nodata		&	\nodata		&	\nodata		&	\nodata 	&	\nodata \\ 
 18.75	&	5 $\pm$ 5	&	\nodata		&	\nodata		&	\nodata		&	\nodata 	&	232 $\pm$ 34 \\ 
 19.25	&	5 $\pm$ 5	&	\nodata		&	\nodata		&	\nodata		&	335 $\pm$ 41	&	521 $\pm$ 51 \\ 
 19.75	&	5 $\pm$ 5	&	\nodata		&	123 $\pm$ 25	&	\nodata 	&	578 $\pm$ 54	&	1043 $\pm$ 73 \\ 
 20.25	&	5 $\pm$ 5	&	154 $\pm$ 28	&	253 $\pm$ 36	&	625 $\pm$ 56	&	1203 $\pm$ 78	&	1839 $\pm$ 97 \\ 
 20.75	&	87 $\pm$ 21	&	346 $\pm$ 42	&	464 $\pm$ 49	&	1198 $\pm$ 78	&	2076 $\pm$ 103	&	2898 $\pm$ 122 \\ 
 21.25	&	191 $\pm$ 31	&	557 $\pm$ 53	&	873 $\pm$ 67	&	2087 $\pm$ 103	&	3104 $\pm$ 126	&	4391 $\pm$ 150 \\ 
 21.75	&	418 $\pm$ 46	&	1079 $\pm$ 74	&	1415 $\pm$ 85	&	2794 $\pm$ 120	&	4990 $\pm$ 160	&	6297 $\pm$ 180 \\ 
 22.25	&	821 $\pm$ 65	&	1932 $\pm$ 99	&	2464 $\pm$ 112	&	4510 $\pm$ 152	&	7501 $\pm$ 196	&	9738 $\pm$ 224 \\ 
 22.75	&	2014 $\pm$ 102	&	3290 $\pm$ 130	&	3952 $\pm$ 142	&	7000 $\pm$ 190	&	10973 $\pm$ 238	&	12688 $\pm$ 256 \\ 
 23.25	&	3699 $\pm$ 138	&	6344 $\pm$ 181	&	6840 $\pm$ 187	&	10823 $\pm$ 236	&	15255 $\pm$ 280	&	18055 $\pm$ 305 \\ 
 23.75	&	7728 $\pm$ 199	&	11784 $\pm$ 246	&	11448 $\pm$ 243	&	16516 $\pm$ 292	&	22617 $\pm$ 341	&	24658 $\pm$ 356 \\ 
 24.25	&	13111 $\pm$ 260	&	19564 $\pm$ 317	&	18918 $\pm$ 312	&	24224 $\pm$ 353	&	32516 $\pm$ 409	&	36705 $\pm$ 435 \\ 
 24.75	&	18737 $\pm$ 311	&	30274 $\pm$ 395	&	28977 $\pm$ 386	&	34618 $\pm$ 422	&	46170 $\pm$ 488	&	46485 $\pm$ 490 \\ 
 25.25	&	28316 $\pm$ 382	&	44956 $\pm$ 481	&	41376 $\pm$ 462	&	46289 $\pm$ 489	&	50778 $\pm$ 512	&	 \\ 
 25.75	&	41386 $\pm$ 462	&	60434 $\pm$ 558	&	43453 $\pm$ 473	&	59370 $\pm$ 553	&			&	 \\ 
 26.25	&	51062 $\pm$ 513	&	69986 $\pm$ 601	&			&			&			&	 \\ 
\enddata
\end{deluxetable}

\clearpage
\begin{deluxetable}{ccccc}
\tablecaption{Number density of $U$ dropouts\label{Udrops}}
\tablehead{
\colhead{$R$ magnitude} & 
\colhead{With Stars} &
\colhead{} &
\colhead{Without Stars} &
\colhead{}\\
\colhead{N arcmin$^{-2}$ 0.5 mag$^{-1}$ } &
\colhead{Error}&
\colhead{N arcmin$^{-2}$ 0.5 mag$^{-1}$ } &
\colhead{Error}
}
\startdata 
 22.25 & 0.017 & 0.005 & 0.003 & 0.002\\
 22.75 & 0.030 & 0.006 & 0.017 & 0.005\\
 23.25 & 0.040 & 0.007 & 0.031 & 0.007\\
 23.75 & 0.095 & 0.012 & 0.095 & 0.012\\
 24.25 & 0.214 & 0.017 & 0.214 & 0.017\\
 24.75 & 0.450 & 0.025 & 0.450 & 0.025\\
 25.25 & 0.661 & 0.031 & 0.661 & 0.031\\
 25.75 & 0.653 & 0.030 & 0.653 & 0.030\\

\enddata
\end{deluxetable}

\clearpage
\begin{deluxetable}{ccc}
\tablecaption{Number density of $B$ dropouts\label{Bdrops}}
\tablehead{
\colhead{$I$ magnitude} & 
\colhead{N arcmin$^{-2}$ 0.5 mag$^{-1}$} &
\colhead{Error}}
\startdata 
 23.25 & 0.0 & \nodata \\
 23.75 & 0.014 & 0.004\\
 24.25 & 0.035 & 0.007\\
 24.75 & 0.099 & 0.011\\
 25.25 & 0.198 & 0.017\\
\enddata
\end{deluxetable}

\clearpage
\begin{deluxetable}{ccc}
\tablecaption{Number density of $V$ dropouts.\label{Vdrops}}
\tablehead{
\colhead{$I$ magnitude \tablenotemark{a}} & 
\colhead{N arcmin$^{-2}$ 0.5 mag$^{-1}$}&
\colhead{Error}}
\startdata 
 23.75 & 0.011 & 0.004\\
 24.25 & 0.019 & 0.005\\
 24.75 & 0.051 & 0.008\\
 25.25 & 0.059 & 0.009\\
 25.75 & 0.061 & 0.009\\
\enddata
\tablenotetext{a}{$I_{c}=I-0.453$ for comparison to \citet{2003PASJ..astroph}}
\end{deluxetable}

\clearpage
\begin{deluxetable}{cccc}
\tablecaption{Number density of $V$ dropouts not detected at 2$\sigma$ in $U$ or $B$ with $V-I > 2.4$ \label{Vdrops_noB}}
\tablehead{
\colhead{$I$ magnitude} & 
\colhead{N arcmin$^{-2}$ 0.5 mag$^{-1}$}&
\colhead{Error}&
\colhead{Percentage of Interlopers}}
\startdata 
 23.75 & 0.001 & 0.001 & 90.9\\
 24.25 & 0.010 & 0.004 & 47.4\\
 24.75 & 0.038 & 0.007 & 25.5\\
 25.25 & 0.047 & 0.008 & 20.3\\
\enddata
\end{deluxetable}

\end{document}